\documentclass[twocolumn,reprint, amsmath,amssymb,showpacs,superscriptaddress]{revtex4-1}

\usepackage[utf8]{inputenc}
\usepackage[T1]{fontenc}
\usepackage[english]{babel}

\usepackage{times}
\usepackage{graphicx}
\graphicspath{{figs/}}
\usepackage{amssymb,amsmath,amsfonts,amsthm}
\usepackage{mathtools}
\usepackage{verbatim}
\usepackage{enumerate}
\usepackage[colorlinks,linkcolor=dred,urlcolor=dblue,citecolor=dgreen]{hyperref}
\usepackage{bbm}
\usepackage{booktabs}

\usepackage{color}
\definecolor{dred}{rgb}{.8,0.2,.2}
\definecolor{ddred}{rgb}{.8,0.5,.5}
\definecolor{dblue}{rgb}{.2,0.2,.8}
\definecolor{dgreen}{rgb}{.2,0.5,.2}

\newcommand{\be}{\begin{equation}}
\newcommand{\ee}{\end{equation}}
\newcommand{\bea}{\begin{eqnarray}}
\newcommand{\eea}{\end{eqnarray}}
\newcommand{\bra}[1]{\mbox{$\langle #1|$}}
\newcommand{\ket}[1]{\ensuremath{|#1\rangle}}

\newcommand{\tr}{\textrm{tr}}

\begin{document}

\def\thesection{%
\Roman{section}}%
\def\thesubsection{%
\Alph{subsection}}%
\def\thesubsubsection{%
\arabic{subsubsection}}%
\def\theparagraph{%
\Roman{paragraph}}%
\def\thesubparagraph{%
\theparagraph.\Roman{subparagraph}}%
\setcounter{secnumdepth}{5}%

\title{Enhancing quantum control by bootstrapping a quantum processor of 12 qubits}

\author{Dawei Lu}
\email{ludw@sustc.edu.cn}
\affiliation{Department of Physics, Southern University of Science and Technology, Shenzhen 518055, China}
\affiliation{Institute for Quantum Computing and Department of Physics and Astronomy,
University of Waterloo, Waterloo N2L 3G1, Ontario, Canada}

\author{Keren Li}
\affiliation{State Key Laboratory of Low-Dimensional Quantum Physics and Department of Physics, Tsinghua University, Beijing 100084, China}
\affiliation{Institute for Quantum Computing and Department of Physics and Astronomy,
University of Waterloo, Waterloo N2L 3G1, Ontario, Canada}

\author{Jun Li}
\affiliation{Beijing Computational Science Research Center, Beijing 100193, China}
\affiliation{Institute for Quantum Computing and Department of Physics and Astronomy,
University of Waterloo, Waterloo N2L 3G1, Ontario, Canada}

\author{Hemant Katiyar}
\affiliation{Institute for Quantum Computing and Department of Physics and Astronomy,
University of Waterloo, Waterloo N2L 3G1, Ontario, Canada}

\author{Annie Jihyun Park}
\affiliation{Institute for Quantum Computing and Department of Physics and Astronomy,
University of Waterloo, Waterloo N2L 3G1, Ontario, Canada}
\affiliation{Max-Planck-Institutf\"{u}r Quantenoptik, D-85748 Garching, Germany}

\author{Guanru Feng}
\affiliation{Institute for Quantum Computing and Department of Physics and Astronomy,
University of Waterloo, Waterloo N2L 3G1, Ontario, Canada}

\author{Tao Xin}
\affiliation{State Key Laboratory of Low-Dimensional Quantum Physics and Department of Physics, Tsinghua University, Beijing 100084, China}
\affiliation{Institute for Quantum Computing and Department of Physics and Astronomy,
University of Waterloo, Waterloo N2L 3G1, Ontario, Canada}

\author{Hang Li}
\affiliation{State Key Laboratory of Low-Dimensional Quantum Physics and Department of Physics, Tsinghua University, Beijing 100084, China}
\affiliation{Institute for Quantum Computing and Department of Physics and Astronomy,
University of Waterloo, Waterloo N2L 3G1, Ontario, Canada}

\author{Guilu Long}
\affiliation{State Key Laboratory of Low-Dimensional Quantum Physics and Department of Physics, Tsinghua University, Beijing 100084, China}

\author{Aharon Brodutch}
\affiliation{Institute for Quantum Computing and Department of Physics and Astronomy,
University of Waterloo, Waterloo N2L 3G1, Ontario, Canada}
\affiliation{Center for Quantum Information and Quantum Control, Department of Physics and Department of Electrical and Computer Engineering, University of Toronto, Toronto M5S 3H6, Ontario, Canada}

\author{Jonathan Baugh}
\affiliation{Institute for Quantum Computing and Department of Physics and Astronomy,
University of Waterloo, Waterloo N2L 3G1, Ontario, Canada}

\author{Bei Zeng}
\email{zengb@uoguelph.ca}
\affiliation{Department of Mathematics and Statistics, University of Guelph, Guelph N1G 2W1, Ontario, Canada}
\affiliation{Institute for Quantum Computing and Department of Physics and Astronomy,
University of Waterloo, Waterloo N2L 3G1, Ontario, Canada}

\author{Raymond Laflamme}
\affiliation{Institute for Quantum Computing and Department of Physics and Astronomy,
University of Waterloo, Waterloo N2L 3G1, Ontario, Canada}
\affiliation{Perimeter Institute for Theoretical Physics, Waterloo N2L 2Y5, Ontario, Canada}

\begin{abstract}
Accurate and efficient control of quantum systems is one of the central challenges for  quantum information processing.  Current state-of-the-art experiments rarely go beyond 10 qubits  and in most cases demonstrate only limited control. Here we demonstrate control of a 12-qubit system, and show that the system can be employed as a quantum processor to optimize its own control sequence by using measurement-based feedback control (MQFC). The final product is a control sequence for a complex 12-qubit task: preparation of a 12-coherent state.  The control sequence is about 10\% more accurate than the one generated by the standard (classical) technique, showing that MQFC can correct for unknown imperfections. Apart from demonstrating a high level of control over a relatively large system, our  results show that even at the 12-qubit level, a quantum processor can be a useful lab instrument. As an extension of our work, we propose a method for combining the MQFC technique with a twirling protocol, to optimize the control sequence that produces a desired Clifford gate.
\end{abstract}
\maketitle

\section{Introduction}

Quantum computers promise to outperform their classical counterparts in many applications \cite{shor1994algorithms,grover1996fast,farhi2001quantum,buluta2009quantum,georgescu2014quantum,harrow2009quantum}.  A primary obstacle in building large-scale quantum computers is the inadequacy of classical computers for the task of optimizing the experimental control field \cite{ladd2010quantum}.  Standard classical optimization algorithms are impractical in the long run since they have a running time that grows exponentially with the number of quantum bits (qubits) \cite{gradl2006parallelising}. In theory, a complex quantum circuit can be decomposed into  elementary gates that work on a restricted number of qubits (usually one or two) and should be readily implemented in experiment \cite{nielsen2010quantum}. In reality however, the control fields are never localized and the qubits interact and evolve  even in the absence of the control  fields. Consequently, the implementation of each elementary gate may require a control sequence that takes into account a subsystem involving many more than one or two qubits. Moreover, the number of elementary gates required for a quantum algorithm grows polynomially with the system size and the errors accumulate with each successive gate. Therefore, an effective and efficient way to optimize the control field and minimize errors is a key ingredient for scaling up quantum information processing devices \cite{brif2010control}.

Here we consider the task of optimizing a control field that will drive the  quantum system from a fixed input state $\rho_i$ to a desired target state $\rho_f$. This problem is important in quantum information processing, as numerous tasks, such as algorithmic cooling in ensemble quantum computing \cite{boykin2002algorithmic,baugh2005experimental}, magic state preparation in fault-tolerant quantum computing \cite{souza2011experimental} and encoding in quantum key distribution \cite{bennett1984quantum}, all rely on steering states regardless of the propagator. The gradient ascent pulse engineering (GRAPE) algorithm \cite{khaneja2005optimal} is the current state-of-the-art algorithm to  (classically) optimize the control field in quantum state engineering problems. It is  widely used in NMR \cite{ryan2008liquid}, electron spin resonance \cite{zhang2011coherent}, nitrogen-vacancy centers in diamond \cite{waldherr2014quantum,dolde2014high}, superconducting circuits \cite{motzoi2009simple,egger2013optimized}, and ion traps \cite{nebendahl2009optimal,schindler2011experimental}. The GRAPE method exploits the gradient of a fidelity function to update the control field iteratively.

GRAPE has two major drawbacks that are indeed common to all classical optimization algorithms: its running time is exponential in the size of the $n$-qubit system, and its accuracy depends on the precision of experimentally obtained  parameters describing the quantum system (e.g., the system Hamiltonian). Basically, it is a gradient-based iterative algorithm. At each iteration $k$, the algorithm computes the evolution of the system under the previous pulse, and produces a final state $\tilde{\rho}$ and a fitness function $f=\tr(\tilde{\rho}\rho_f)$. It then computes the current gradient $g$ for the use of updating the pulse. Classically, the computation involves the matrix exponential and multiplication in the  %$2^n\times2^n$
$2^n$-
dimensional  Hilbert space and hence takes an exponential (in the number of qubits $n$) amount of time. For instance, a cluster of 128 AMD \emph{Opteron 850} CPU (2.4 GHz) can only handle a problem size of about ten qubits using GRAPE \cite{gradl2006parallelising}.

\begin{figure*}
\center
\includegraphics[width=1.8\columnwidth]{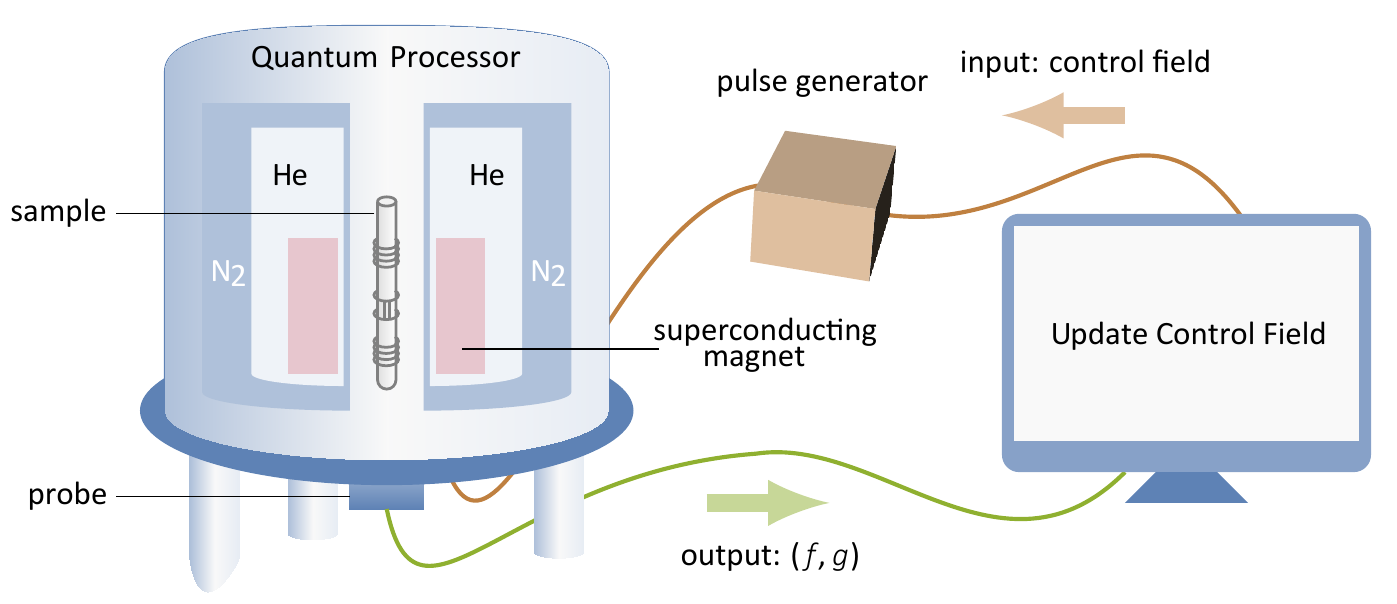}
\caption{MQFC process for optimizing a control field. Starting from an initial guess, a shaped pulse is created from the pulse generator and then applied to the sample. The fidelity function $f$ of the control pulse and its gradient $g$ are directly measured on the quantum processor, where $g$ is used for   updating the control field till that sufficiently high fidelity $f$ has been   achieved.}
\label{fig1}
\end{figure*}

%it is proposed that a quantum processor can be used to calculate $f$ and $g$ efficiently \cite{li2016physical,rebentrost2016quantum}.
Recently, Li et al. \cite{li2016physical} and later Rebentrost et al.  \cite{rebentrost2016quantum}   showed that a quantum processor can be used to   calculate $f$ and $g$ efficiently.   A technique called  measurement-based quantum feedback control (MQFC)  enables direct measurement of $f$ and $g$ (see  Fig. \ref{fig1})%.   MQFC uses
, allowing the quantum processor to optimize its own pulses. MQFC addresses both the issues of scalability and control inaccuracies due to imperfect system characterization \cite{vijay2012stabilizing,hirose2016coherent}. Moreover, this technique is transferrable to any implementation in which control fields steer the system evolution and measurement in a standard basis is possible. In this work, we implement MQFC on a 12-qubit NMR quantum processor, and in particular demonstrate for the first time that MQFC  enhances the control precision by about 10\% due to its self-feedback property. Furthermore, by creating  the 12-coherent state  %and its direct observation using a single probe qubit,
we demonstrate the  capability of our quantum processor to  function  as a universal 12-qubit quantum processor with high-fidelity individual controls. This is also one of the largest quantum processors with individual-control to date.

\section{Results}

In this paper, we refer to unnormalized deviation density matrices (without the identity term)  as `states', which is a standard convention in ensemble quantum computing. To distinguish from the Hamiltonian, we use capital X, Y, and Z to denote states and $\sigma_x$, $\sigma_y$ and $\sigma_z$ to denote Hamiltonians, while they both refer to the same set of Pauli matrices.

\begin{figure*}
\includegraphics[width=0.9\linewidth]{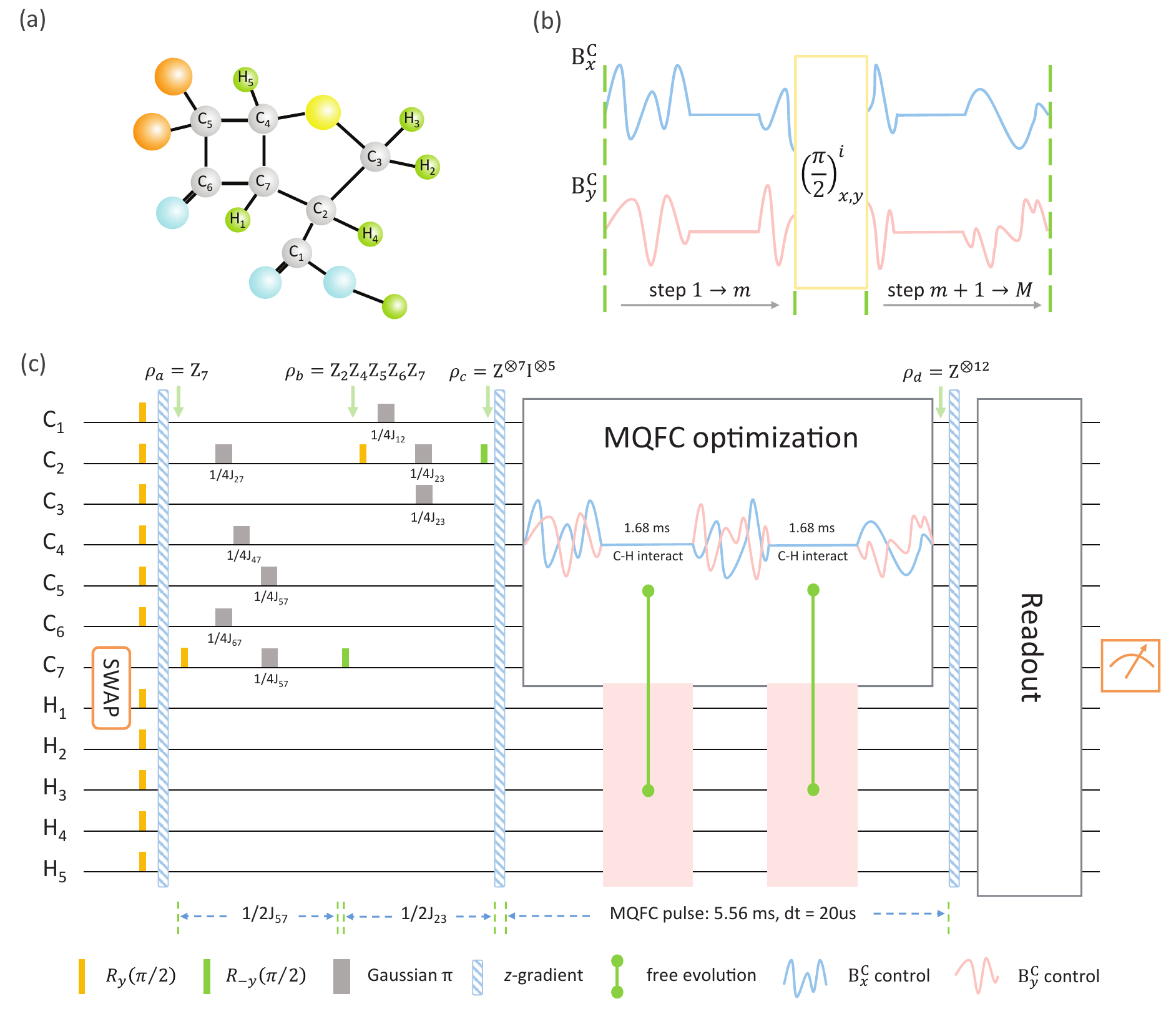}
\caption{MQFC scheme in creating 12-coherence. \textbf{a} Molecular structure of the 12-qubit quantum processor. \textbf{b} Schematic of measuring the $m$-th step gradient $g_{x,y}[m]$. A $\pi/2$ rotation about $x (y)$-axis for qubit $i$ is inserted between the $m$-th and $(m+1)$-th slices. \textbf{c} Quantum circuit that evolves the system from the thermal equilibrium to 12-coherence, where MQFC is applied on 7-coherence Z$^{\otimes 7}$I$^{\otimes 5}$. }
\label{fig2}
\end{figure*}

\emph{Quantum processor.} -- In our NMR quantum processor, the liquid-state sample is per-$^{13}$C labeled (1S,4S,5S)-7,7-dichloro-6-oxo-2-thiabicyclo[3.2.0]heptane-4-carboxylic acid dissolved in d6-acetone, which forms a 12-qubit register. The 12 qubits are denoted by nuclear spins C$_1$ to C$_7$ ($^{13}$C-labeled) as qubits 1 to 7, and H$_1$ to H$_5$ as qubits 8 to 12 in the molecule shown by Fig. \ref{fig2}a. When placed in a static $z$-magnetic field, it has a system Hamiltonian
\be
\mathcal{H}_{s} =  -\pi\sum_{i=1}^{12} \nu_0^i \sigma_z^{i} + \frac{\pi}{2}\sum_{i=1<j}^{12} J_{ij} \sigma_z^{i}\sigma_z^{j},
\label{Hnmr}
\ee
where $\nu_0^i$ is the Larmor frequency of the $i$th qubit, $J_{ij}$ is the coupling between qubits $i$ and $j$, and $\sigma_z^i$ is the Pauli-z operator of the $i$th qubit. The values of these parameters can be found in Appendix C \cite{supple}.

The control Hamiltonian is due to the transverse control field applied in the $x$-$y$ plane, which is often digitized into $M$ slices with slice length $\Delta t$. In each slice, there are four constant control parameters, leading to a control Hamiltonian in the form of
\begin{align}
\mathcal{H}_{c}[m]  = {} & \text{B}_x^\text{C}[m]\sum_{i=1}^7 \sigma_x^i + \text{B}_y^\text{C}[m]\sum_{i=1}^7 \sigma_y^i  \\ \nonumber
+{}  &  \text{B}_x^\text{H}[m]\sum_{j=8}^{12} \sigma_x^j + \text{B}_y^\text{H}[m]\sum_{j=8}^{12} \sigma_y^j,
\label{Hext}
\end{align}
where, for example, $\text{B}_x^\text{C}[m]$ means the $x$-component of the $m$th slice of control field     in the $^{13}$C channel.

The dynamics of the NMR system is governed by $\mathcal{H}_{s}$ and $\mathcal{H}_{c}$ simultaneously, with the propagator
\be
U_{1}^{M} = U_MU_{M-1}\cdots U_{1},
\label{Um}
\ee
where
\be
U_m = e^{-i(\mathcal{H}_{s}+\mathcal{H}_{c}[m])\Delta t}.
\ee
The essence of NMR quantum information processing is to optimize a control field, i.e. find a sequence of $\text{B}_{x,y}[m]$, such that one can precisely realize a quantum gate or drive the system to a target state according to Eq. (\ref{Um}).

\emph{Fundamentals of the GRAPE algorithm.} -- To implement  a particular target gate or state we  need to find an  optimal $\text{B}_{x,y}[m]$.  One of the most prominent optimization algorithms to date is the GRAPE algorithm \cite{khaneja2005optimal} which was developed for the design of optimal control pulses in NMR spectroscopy.
Here, we explain the basic principle of   GRAPE   by considering the problem of state engineering in the absence  of relaxation.

Suppose the initial  state of the spin system is $\rho_i$, and the target output state is $\rho_f$. After applying a $M$-slice trial control pulse,  the system will evolve to
\begin{equation}
\tilde \rho = U_{1}^{M} \left(\rho_i\right) = U_{1}^{M} \rho_i U_{1}^{M\dag}.
\label{finalstate}
\end{equation}
The fitness function defined as $f = \operatorname{tr} (\rho_f \tilde \rho)$ serves as a metric for the control fidelity, with the form
 \be
 f = \operatorname{tr} (\rho_f \tilde \rho) = \operatorname{tr} \left( U_{1}^{M} \left(\rho_i\right) \cdot \rho_f \right).
 \label{fitness}
 \ee
Obviously, $f$ is a function of $2M$ variables, and to find its optimium we calculate the gradient function to the first order
\bea
&& g_{x,y} [m]   =  \frac{\partial f}{\partial \text{B}_{x,y}[m]}    \nonumber \\  \label{gradient2}
 && \approx  \sum\limits_{k = 1}^n {\operatorname{tr} \left( { - i \Delta t \cdot {U_{m+1}^{M}} \left[ {\sigma_{x,y}^k,U_{1}^{m} \left(\rho_i\right)}  \right] U_{m+1}^{M\dag} \cdot  \rho_f} \right) },
\eea
where $\left[ {\sigma_{x,y}^k,U_{1}^{m} \left(\rho_i\right)}  \right]$ is the commutator between $\sigma_{x,y}^k$ and $U_{1}^{m} \left(\rho_i\right)$. We may increase the fitness function $f$ by using the gradient iteration rule
\begin{equation}
\text{B}_{x,y}[m] \leftarrow \text{B}_{x,y}[m] + \epsilon \cdot g_{x,y} [m],
\label{iteration}
\end{equation}
where $\epsilon$ is a suitably chosen step size.

The GRAPE algorithm proceeds as follows on a classical computer:

\emph{1}. start from an initial guess control $\text{B}_{x,y}[m]$;

\emph{2}. calculate $\tilde \rho$ according to Eq. (\ref{finalstate});

\emph{3}. evaluate fitness function $f = \operatorname{tr} (\rho_f \tilde \rho)$;

\emph{4}. if $f$ does not reach our preset value, evaluate gradient function $g$ according to Eq. (\ref{gradient2});

\emph{5}. update control variables according to Eq. (\ref{iteration}), then go to step \emph{2}.

\emph{MQFC optimization.} -- The GRAPE algorithm requires the calculation of $U_{1}^{M}$, i.e., the dynamics of the system. This step is inefficient on a classical computer when the size of the system is large. In contrast, the scheme of MQFC  optimization provides an alternative way which enables direct measurement of $f$ and $g$ in the experimental manner, or explicitly, via the quantum evolution and measurement of the quantum processor.

Without loss of generality, let us discuss the scenario of ensemble quantum computing. e.g., NMR quantum computing, where the state is usually written as a traceless deviation density matrix and a single-shot measurement is sufficient to get the expected value of an observable. For other systems that use the computational basis or projective measurement, the following procedure needs to be slightly modified and more repetitions may be required to get the estimate of $f$ and $g$.

Measuring $f$ is straightforward. For an $n$-qubit system, the total number of elements in the Pauli basis is $4^n-1$ (without the identity term). If the target state $\rho_f$ has some decomposition, say, $\rho_f = \sum_{\gamma=1}^{\mathcal{G}} x_\gamma P_\gamma$ with respect to the Pauli basis, then the fitness function is
\begin{equation}
f = \operatorname{tr}\left( \tilde \rho  {\rho _f} \right)  = \sum\limits_{\gamma=1}^{\mathcal{G}} {{x_r}\operatorname{tr}\left( {\tilde\rho  {P_r}} \right)}. \label{f}
\end{equation}
Here, $1\leq \mathcal{G}\leq 4^n$ denotes the number of nonzero components, $P_\gamma$ is the $\gamma$-th element of the Pauli basis, and $x_\gamma$ is its corresponding coefficient.

 Therefore, $\mathcal{G}$ experiments are required to estimate $f$.  In the $\gamma$-th experiment, we just need to apply the control field to the initial state $\rho_i$ and measure the expectation value $\langle P_\gamma \rangle$ of $\tilde \rho$.  For a generic $\rho_f$ that contains all $\mathcal{G} = 4^n-1$ Pauli terms, measuring $f$ in experiment is equivalent to carrying out full state tomography, and is thus inefficient. However, many tasks require the creation of a \emph{simple} target state where $\mathcal{G}$ is quite small. For instance, if we aim to prepare the 12-coherent state $\rho_f = \text{Z}^{\otimes 12}$, one measurement is sufficient to obtain $f$.

Measuring $g$ requires us to  realize the commutator $[\sigma^k_{x,y}, \cdot]$ inside Eq. (\ref{gradient2}). In fact \cite{li2016physical},
\begin{equation}
\left[ \sigma_{x,y}^k,\rho  \right] =  i \left(\mathcal{R}_{x,y}^{k} \left(\rho \right)   - \overline{\mathcal{R}}_{x,y}^{k} \left(\rho \right)  \right),
\label{commutator}
\end{equation}
in which $\mathcal{R}_{x,y}^{k}$ and $\overline{\mathcal{R}}_{x,y}^{k}$ mean a $\pi/2$ rotation and $-\pi/2$ about $x$ or $y$ axis on the $k$-th qubit, repsectively. By substituting Eq. (\ref{commutator}) into Eq. (\ref{gradient2}), we get
\bea
g_{x,y}[m]  & = &    \Delta t \sum\limits_{k = 1}^n  \operatorname{tr}   \left\{  \left(U_{m+1}^{M}   \mathcal{R}_{x,y}^{k} U_{1}^{m}\right)  (\rho _i) \cdot \rho_f \right\}  \nonumber \\
&-&\Delta t  \sum\limits_{k = 1}^n  \operatorname{tr} \left\{ \left(U_{m+1}^{M}  \overline{\mathcal{R}}_{x,y}^{k} U_{1}^{m}\right) (\rho _i)      \cdot \rho_f \right\}.
\label{g}
\eea
The terms on the right-hand side are very similar to the measurement of $f$ in Eq. (\ref{fitness}), and the only difference is the local $\pm \pi/2$ pulse inserted between slices $m$ and $m+1$. Explicitly, the $m$-th component of $g_{x,y}$ is a weighted sum of $4n\mathcal{G}$ measurement quantities, where $4$ comes from the $\pm \pi/2$ pulses about the $x$ and $y$ axes, $n$ from the sum over all the qubits, and $\mathcal{G}$ from the measurement of $f$. In each experiment, compared to the way of measuring $f$, we just need to insert a local $ \pi/2 $ pulse after the $m$-th slice evolution.  Provided that all the qubits are well individually addressed, high fidelities are attainable in implementing these local $ \pi/2 $ rotations.

In summary, we need $4n \mathcal{G}M$ experiments in total to perform the gradient measurement, which is linear in  the number of qubits.

\emph{Experimental MQFC optimization.} -- Now we turn to the experiment where the MQFC optimization is used to create the 12-coherent state in the 12-qubit quantum processor. First, let us clarify that all other pulses except the MQFC pulse throughout our experiments are local rotations, which are generated from a subsystem-based gradient ascent pulse engineering (SSGRAPE) approach \cite{ryan2008liquid}. It is a technical improvement of the original GRAPE for our particular implementation, but does not address its poor scalability issue (see Appendix D \cite{supple}). What makes the MQFC scheme remarkable is that, it does not involve the computationally expensive classical   simulation of the $2^{12}$-dimensional quantum dynamics  in the course of optimization.

For our optimization task, GRAPE is a powerful tool, but handling 12 qubits is near the limit of capability for a typical laptop computer. In contrast, MQFC is capable of overcoming this difficulty in certain cases. Taking our experiment as an example, MQFC is able to solve the problem of finding a control field that evolves single-coherence ZI$^{\otimes 11}$ into 12-coherence Z$^{\otimes 12}$ in a time that scales linearly with the number of qubits. The entire experimental procedure is depicted in Fig. \ref{fig2}c, with a step-by-step description in Appendix E \cite{supple}.

First, we prepare 7-coherence Z$^{\otimes 7}$I$^{\otimes 5}$ on the seven $^{13}$C spins, using the sequence in Fig. \ref{fig2}c before the \emph{MQFC optimization} box. This procedure, benchmarked in our previous work \cite{lu2015experimental}, is mainly done with the aid of SSGRAPE. Subsequently, we create Z$^{\otimes 12}$ via MQFC on the quantum processor, which is the main focus of this work. We attempt to optimize a control field, namely a shaped radio frequency (r.f.) pulse, to evolve the system from the input $\rho_{i} = \text{Z}^{\otimes 7}\text{I}^{\otimes 5}$ to the output $\rho_{f} = \text{Z}^{\otimes 12}$. Our control field, as shown in the \emph{MQFC optimization} box, is comprised of three sub-pulses to realize local rotations, and two free evolutions to let $^{13}$C qubits interact with $^{1}$H qubits for the purpose of generating higher coherence. The whole control field is digitized into $M=278$ slices with $\Delta t=20$ $\mu$s width, while 110 slices are for three sub-pulses and 168 slices remain zero to realize the two 1.68 ms free evolutions (Appendix E \cite{supple}). The total dynamics of the pulse is given by $U_{1}^{M}$ in Eq. (\ref{Um}).

The fitness function is defined as $f = \tr (\rho_{f} \tilde{\rho})$, a metric for the control fidelity, where $\tilde{\rho}=U_{1}^{M}(\rho_{i})$ is the experimental state and $\rho_{f}= \text{Z}^{\otimes 12}$ is the target. In our experiment, only one measurement of the expectation value of $\langle \text{Z}^{\otimes 12} \rangle$ suffices to attain $f$ after each iteration. If $f$ does not hit our preset value with the current control field, we navigate the control field along its gradient $g$. In fact, to measure $g_x[m]$ (the same for $g_y[m]$) which is the gradient of slice $m$, we just need three steps: insert a local $\pm\pi/2$ pulse on every qubit about $x$-axis between slice $m$ and $m+1$; apply this new control field to the initial state $\rho_i$ and measure $f$ (see Fig. \ref{fig2}b); compute $g_x[m]$ by directly combining these $\pm\pi/2$-inserted results via Eq. (\ref{g}). As long as accurate local $\pm\pi/2$ pulses are available for each qubit, $g$ can be measured on a quantum processor. In experiment, we have designed a 1 ms $\pi/2$ pulse on every $^{13}$C nucleus with the simulated fidelity over $99.7\%$ (Appendix D \cite{supple}). Having the gradient, we can update the control field and continue the MQFC procedure until a desired $f$ is attained.

\emph{ Direct observation of 12-coherence.} -- After the preparation of the 12-coherent state, the next step is to observe it. In NMR spectroscopy, multiple coherence is hard to be observed directly in a one-dimensional spectrum, i.e., by flipping the target spin to the $x$-$y$ plane while others remain in Z. If all coupling between the target spin and other spins can be resolved, such observation is feasible. For example, in a two-qubit system, we can flip spin one to X to observe ZZ. In fact, XZ can be written as
\be
\text{XZ} = \text{X}\otimes \ket{0}\bra{0} - \text{X}\otimes \ket{1}\bra{1}.
\ee
The first term $\text{X} \otimes \ket{0}\bra{0}$ leads to a positive peak at $\nu_1-J_{12}/2$ in the spectrum, as the $J$-coupling term shifts the frequency of qubit 1 by $-J_{12}/2$. Analogously, the second term $\text{X}\otimes \ket{1}\bra{1}$ leads to a negative (due to the minus sign before the term) peak at $\nu_1+J_{12}/2$. Generally, these two peaks can be resolved in the spectrum as long as $J$ is large enough to separate them in frequencies. However, to observe multiple coherence, this requirement is of great challenge, since all $J$-couplings between the target spin and other spins should be sufficiently large to prevent the annihilations of positive and negative peaks. As a result, two-dimensional spectra and special techniques are usually employed to observe multiple coherence in conventional NMR spectroscopy.

\begin{figure*}
\includegraphics[width=0.9\linewidth]{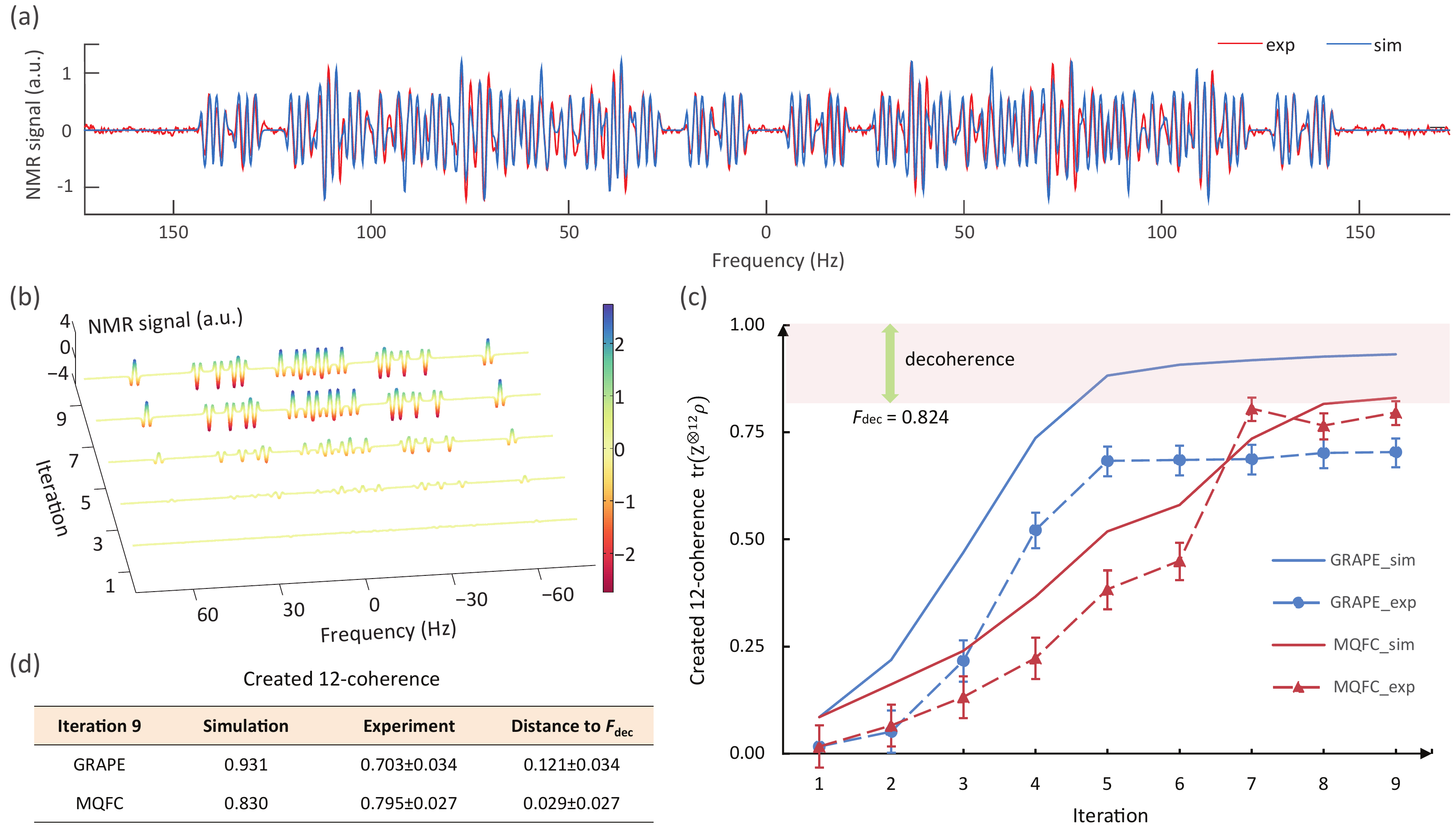}
\caption{Experimentally created 12-coherence using MQFC. \textbf{a} Direct observation of the created 12-coherence in one-dimensional NMR spectrum (red), where C$_7$ is the probe qubit. Simulated spectrum (blue) is also plotted. The experimental result is rescaled by 1.21 times to compensate for the decoherence effect for better visualization. \textbf{b} Spectra of 12-coherence after each odd iteration during the MQFC optimization. Unlike the direct observation, a readout technique is applied to gain a higher resolution. A color scale indicates peak intensities. The height of the peaks is proportional to the value of created 12-coherence. \textbf{c} Comparison between GRAPE (blue) and MQFC (red) optimizations, both in simulation (solid; without decoherence accounted) and experiment (dashed). $F_{dec}$ is the numerical simulation of decoherence during the 12-coherence creation. Compared to the GRAPE algorithm, MQFC optimization is worse in simulation, but better in experiment. The error bars are plotted by the infidelity of the readout pulse. \textbf{d} Results at iteration 9. The experimental 12-coherence reaches $0.795$ using MQFC which approaches the $F_{dec} = 0.824$ bound, while GRAPE only leads to $0.703$ (i.e., $0.121$ lower than $F_{dec}$) in experiment. }
\label{fig3}
\end{figure*}

For the purpose of NMR quantum computing, it is certainly better if one can read out multiple coherence directly in a one-dimensional spectrum, as one-dimensional spectrum reflects the state information more intuitively and reduces experimental running time remarkably compared to the two-dimensional spectroscopy. In our 12-qubit processor, although there are a few couplings as small as 0.01 Hz (Appendix C \cite{supple}), a direct observation of 12-coherence $\text{Z}^{\otimes 12}$ is still available on C$_7$. Figure \ref{fig3}a exhibits a strong agreement between experimental observation 12-coherence with merely 32 scans and the simulation, after rescaling the experimental result by 1.21 times to compensate for decoherence. To our best knowledge, our experiment is the first direct observation of multiple coherence beyond ten spins, and provides a valid evidence that our 12-qubit processor possesses excellent individual controllability and the potential to be a universal 12-qubit quantum processor.

\emph{Readout sequence.} -- Although the direct observation of 12-coherence with 32 scans in Fig. \ref{fig3}a demonstrates our control precision, it is not suitable for the many experimental runs during the optimization since 32 scans leads to a great time cost. One solution is to decouple the five $^1$H spins to boost the signal-to-noise ratio (SNR) by $2^5=32$ times, which exactly compensates for the required scan number. We have designed a readout pulse sequence to realize it as shown in Fig. \ref{readout}.

 The local pulses in the readout sequence are computed by SSGRAPE, and the sequence is implemented before every measurement. The phase correction is a $z$-rotation to neutralize the unwanted chemical shift rotation during the free evolution. If the state is $\text{Z}^{\otimes 12}$, the five $^1$H spins will be evolved to the identity state after the readout sequence, and the decoupling of $^1$H  leads to the C$_7$ spectrum as shown in Fig. \ref{fig3}b, which is measured with a single scan. We then use spectrum fitting to obtain the signal's amplitude and phase, and thus the value of $\langle \text{Z}^{\otimes 12} \rangle$.

This readout sequence induces errors in terms of decoherence and pulse imperfections. For the former one, through our simulation we find that it leads to about $30\%$ signal loss, which is reasonable since multi-coherence is exceptionally vulnerable to decoherence. Therefore, this factor is taken into account for all the measurement results, that is, the measured values are rescaled by about 1.3. With respect to the pulse imperfection, it consists of two parts: the imperfection of the sequence itself, i.e., some approximations when we design this simple readout sequence, and the infidelities in implementing the pulses. In total, 3.5\% error arises in simulation. We use this value as the uncertainty of the experimental value of $\langle \text{Z}^{\otimes 12} \rangle$, namely, the error bars in Fig. \ref{fig3}c.

\emph{Experimental results.} -- Figure \ref{fig3}b shows the spectrum of $\tilde{\rho}$ after the readout stage for each odd iteration. The peak intensities correspond to the value of $f = \tr(\text{Z}^{\otimes 12}\tilde{\rho})$, which clearly shows that MQFC increases $f$ during the optimization. This demonstrates that MQFC is a practical technique for designing control fields in large quantum systems.

Our experiment also exhibits MQFC's ability of correcting unknown experimental errors. To demonstrate this improvement, we implement another group of 12-coherence-creating experiments, where all experimental settings are the same except that the pulse is generated from the classical SSGRAPE method other than the MQFC approach. We then compare these two groups of experiments. Figure \ref{fig3}c illustrates the result of SSGRAPE and MQFC pulses both in simulation and experiment. Focusing on the final result at iteration 9 in Fig. \ref{fig3}d, in experiment SSGRAPE finally creates a 12-coherence with $f=0.703\pm 0.034$, whereas MQFC pulse creates $f=0.795 \pm 0.027$. This experimental improvement (nearly $10\%$) disagrees with simulation, as in simulation MQFC (0.830) is even worse than SSGRAPE (0.931).

Considering that MQFC is a feedback-control process, some incomplete knowledge of the experimental quantum process, such as the nonlinearity of the pulse generator or imprecision of the molecular Hamiltonian, may be inherently corrected during the optimization. Indeed, the experiment clearly suggests that MQFC is advantageous in terms of correcting errors from unknown sources. Furthermore, we simulate the decoherence effect during the procedure, and find that the upper bound of $\tr(\text{Z}^{\otimes 12}\tilde{\rho})$ in the presence of dephasing noise is about $0.824$ (see Methods). Note that our MQFC result finally reaches $0.795$, which is very close to this bound, demonstrating that our control of this 12-qubit processor is close to the theoretical prediction after accounting for decoherence.

\section{Discussion}
\label{sec4}

\emph{Scalability.} -- One major concern about control methods is their scalability with the number of qubits $n$.   Our MQFC protocol involves a single experiment to measure $f$ and $4nM$ experiments to measure $g$ for each iteration, where $n$ is the number of qubits. Assuming each experiment takes $\tau_{exp}$ time, the MQFC in total consumes $T_{it}=  (4nM+1)\tau_{exp}$ for each iteration. For comparison, one has to deal with massive $2^n\times2^n$ matrix multiplications and exponentials using GRAPE on a classical computer. The speed-up comes from the fact that MQFC utilizes the evolution of the quantum system instead of computing the system's dynamics when evaluating $f$ and $g$.

For  other potential problems when scaling up the GRAPE technique, MQFC confronts similar difficulties, such as how to effectively represent a generic target state, how to choose a good initial guess, how to determine the pulse parameters before optimization, and how many iterations are needed to reach a satisfactory fidelity. Unfortunately, experimental observation of  running time versus number of qubits is not likely in NMR, since changing the number of qubits would usually require a different sample with different characteristics. So we cannot experimentally compare the scaling of MQFC versus GRAPE, instead we must be satisfied with the fact that MQFC performs well at the 12-qubit level and should theoretically scale better than GRAPE under standard assumptions. See Appendix A in \cite{supple} for details.

One may also ask if there could be other classical algorithms that scale as well (or better than) MQFC. This question remains open, but it seems very unlikely -- the gradient calculation is based on the dynamics as shown in Eq. (\ref{Um}), i.e., the expected classical algorithm needs to simulate the dynamics of an NMR system in an efficient way. Even when boiling down to our particular state engineering task, as far as we can tell, there is no employed numerical method \cite{daley2014quantum,mascarenhas2015matrix,lloyd2014information} to simplify such an optimization, despite extensive work on the subject since the early days of experimental quantum computing. Moreover, MQFC can correct unknown errors to some extent, while   open-loop algorithms should require knowledge about   the noise spectrum in advance, which is usually  impractical for large quantum systems. In this sense, another potential application of MQFC is to demonstrate the quantum computing supremacy \cite{preskill2012quantum}, where initial endeavors have been made in other systems, for example in a recent  five-photon boson sampling experiment  \cite{wang2017high}.

\emph{Optimizing Clifford gates.} -- While our experiment focuses on state engineering,  MQFC can also be used for other quantum optimization tasks.  As an example, we consider optimizing the pulse sequence for a generic Clifford gate.  It is possible to use twirling to estimate the average gate fidelity of a Clifford gate efficiently  \cite{lu2015experimental}.  The twirling protocol is based on finding the fidelity between  experimental states following the pulse sequence and the corresponding desired states following the ideal gate.  In principle this should be done for a complete set of initial states, but a randomized protocol can be used to approximate the gate fidelity with a constant number of experiments.  The MQFC protocol can be modified to extract the desired fidelities and optimize the pulse sequence accordingly (details in Appendix B in \cite{supple}).  Note that right after our work, a five-qubit implementation of a different quantum algorithm for gate optimization was reported \cite{dive2017situ}.

\emph{Comparison with previous work.} -- MQFC was originally introduced in Ref. \cite{li2016physical} where it was implemented on a   7-qubit NMR processor.  There are two significant improvements in our work. First, our work clearly demonstrates the superiority of MQFC in correcting unknown errors with around 10\% fidelity boost compared to the best classical optimization result, while in the 7-qubit experiment no improvement was observed. The reason could be that the characterization of a 7-qubit system is much more accurate than a 12-qubit one, indicating that MQFC should be more powerful when dealing with large systems as the knowledge of larger systems are more likely to be incomplete. Second, our 12-qubit experiment lies at the cutting edge of  present experimental quantum computing, and the capability of individual controls at this qubit number is state-of-the-art. As a comparison, in a recent work \cite{song201710}, the 10-qubit entanglement in a superconducting circuit is created with fidelity 0.668 using \emph{global} control. Moreover, we demonstrated that at the 12-qubit level, the algorithm is already fast enough to justify its use as a tool in the lab.

In summary, we have created a 12-coherence state on an NMR quantum processor using MQFC. Our experimental procedure and result, in particular the direct observation of 12-coherence with one qubit as the probe, signify the capability of our quantum processor to serve as a universal 12-qubit quantum processor with high-fidelity individual controls on each qubit. In terms of control field optimization, our experiment demonstrates two superiorities in efficiency and experimental performance of MQFC beyond its classical counterpart. MQFC requires a running time that scales linearly with the number of qubits, and yields about $10\%$ improvement compared to the best result via classical optimization. This optimization approach could be exceptionally useful in a large system with incomplete characterization, and is readily transferrable to other systems such as superconducing circuits or nitrogen-vacancy centers in diamond. We expect that, as experiments involving more than 10 qubits become more common, quantum feedback methods such as MQFC will become standard tools in quantum computing labs.

\section{Methods}

To numerically simulate the decoherence effect in our 12-qubit system, we first make the following assumptions: the environment is Markovian; only the $T_2^*$ dephasing mechanism is taken into account since $T_1$ effect is negligible in our circuit; the dephasing noise is independent between all qubits; the dissipator and the total Hamiltonian commute in each pulse slice as $\Delta t = 20$ $\mu$s is small. With these assumptions, we solve the master equation in two steps for each $\Delta t$: evolve the system by the propagator in Eq. (\ref{Um}), and subsequently apply the dephasing noise for $\Delta t$ which is an exponential decay of off-diagonal elements in the density matrix. The typical length of simulating our 12-qubit experiment in the presence of dephasing noise is in the magnitude of days on a desktop computer. The simulation shows that at most $F_{dec}=0.824$ of $\text{Z}^{\otimes 12}$ can be achieved with the 5.56 ms MQFC pulse applied on Z$^{\otimes 7}$I$^{\otimes 5}$, which is reasonable as high-order coherence is very vulnerable to the dephasing noise. Alternatively speaking, the upper bound of the MQFC experimental result is 0.824, since the optimization procedure does not include the function of robustness against dephasing noise yet.

\textbf{Data availability}

The data sets generated during and/or analyzed during the current study are available from the corresponding author on reasonable request.

\textbf{Supplementary Information} is available in the online version of the paper.

\textbf{Acknowledgements.} We thank Anthony P. Krismanich, Ahmad Ghavami, and Gary I. Dmitrienko for synthesising the NMR sample.  This
research was supported by CIFAR, NSERC and Industry of Canada. K.L., H.L. and G.L. acknowledge National Natural Science Foundation of China under Grants No. 11175094 and No. 91221205.

\textbf{Competing Interests.} The authors declare that they have no
competing financial interests.

\textbf{Author Contributions.} D.L., J.L. and R.L. conceived the experiments. D.L. and K.L. performed the experiment and analysed the data. D.L., H.K., A.P., G.F. and H.L. tested the sample and developed the control techniques. J.L. and A.B. provided theoretical support. G.L., J.B., B.Z. and R.L. supervised the project. D.L., K.L. and J.L. wrote the manuscript with feedback from all authors. D.L., K.L., and J.L. contributed equally to this work.

\newpage

\newpage

\newpage

\newpage
\begin{figure*}[t]
\begin{center}
\includegraphics[width=\linewidth]{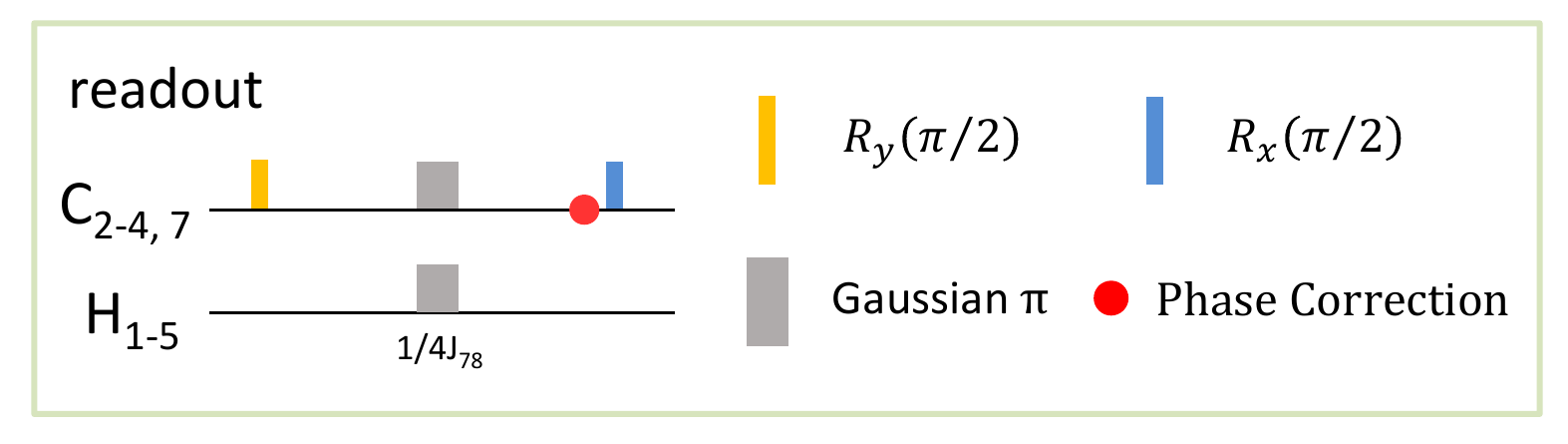}
\end{center}
\caption{Readout sequence to boost the SNR of the C$_7$ spectrum. It transforms the $^1$H spins from Z to identity and thus enables the decoupling of $^1$H channel. The phase correction compensates for the chemical shift evolutions, after which all relevant spins are along the $y$-axis.  In principle, this technique improves the SNR by a factor of 32, and makes the measurement of $f$ or $g$ practical using one scan.}\label{readout}
\end{figure*}

\clearpage

\title{Supplemental Information: Enhancing quantum control by bootstrapping a quantum processor of 12 qubits}

\author{Dawei Lu}
\email{ludw@sustc.edu.cn}
\affiliation{Department of Physics, Southern University of Science and Technology, Shenzhen 518055, China}
\affiliation{Institute for Quantum Computing and Department of Physics and Astronomy,
University of Waterloo, Waterloo N2L 3G1, Ontario, Canada}

\author{Keren Li}
\affiliation{State Key Laboratory of Low-Dimensional Quantum Physics and Department of Physics, Tsinghua University, Beijing 100084, China}
\affiliation{Institute for Quantum Computing and Department of Physics and Astronomy,
University of Waterloo, Waterloo N2L 3G1, Ontario, Canada}

\author{Jun Li}
\affiliation{Beijing Computational Science Research Center, Beijing 100193, China}
\affiliation{Institute for Quantum Computing and Department of Physics and Astronomy,
University of Waterloo, Waterloo N2L 3G1, Ontario, Canada}

\author{Hemant Katiyar}
\affiliation{Institute for Quantum Computing and Department of Physics and Astronomy,
University of Waterloo, Waterloo N2L 3G1, Ontario, Canada}

\author{Annie Jihyun Park}
\affiliation{Institute for Quantum Computing and Department of Physics and Astronomy,
University of Waterloo, Waterloo N2L 3G1, Ontario, Canada}
\affiliation{Max-Planck-Institutf\"{u}r Quantenoptik, D-85748 Garching, Germany}

\author{Guanru Feng}
\affiliation{Institute for Quantum Computing and Department of Physics and Astronomy,
University of Waterloo, Waterloo N2L 3G1, Ontario, Canada}

\author{Tao Xin}
\affiliation{State Key Laboratory of Low-Dimensional Quantum Physics and Department of Physics, Tsinghua University, Beijing 100084, China}
\affiliation{Institute for Quantum Computing and Department of Physics and Astronomy,
University of Waterloo, Waterloo N2L 3G1, Ontario, Canada}

\author{Hang Li}
\affiliation{State Key Laboratory of Low-Dimensional Quantum Physics and Department of Physics, Tsinghua University, Beijing 100084, China}
\affiliation{Institute for Quantum Computing and Department of Physics and Astronomy,
University of Waterloo, Waterloo N2L 3G1, Ontario, Canada}

\author{Guilu Long}
\affiliation{State Key Laboratory of Low-Dimensional Quantum Physics and Department of Physics, Tsinghua University, Beijing 100084, China}

\author{Aharon Brodutch}
\affiliation{Institute for Quantum Computing and Department of Physics and Astronomy,
University of Waterloo, Waterloo N2L 3G1, Ontario, Canada}
\affiliation{Center for Quantum Information and Quantum Control, Department of Physics and Department of Electrical and Computer Engineering, University of Toronto, Toronto M5S 3H6, Ontario, Canada}

\author{Jonathan Baugh}
\affiliation{Institute for Quantum Computing and Department of Physics and Astronomy,
University of Waterloo, Waterloo N2L 3G1, Ontario, Canada}

\author{Bei Zeng}
\email{zengb@uoguelph.ca}
\affiliation{Department of Mathematics and Statistics, University of Guelph, Guelph N1G 2W1, Ontario, Canada}
\affiliation{Institute for Quantum Computing and Department of Physics and Astronomy,
University of Waterloo, Waterloo N2L 3G1, Ontario, Canada}

\author{Raymond Laflamme}
\affiliation{Institute for Quantum Computing and Department of Physics and Astronomy,
University of Waterloo, Waterloo N2L 3G1, Ontario, Canada}
\affiliation{Perimeter Institute for Theoretical Physics, Waterloo N2L 2Y5, Ontario, Canada}

\maketitle

\begin{figure*}
\begin{center}
\includegraphics[width= 0.7\linewidth]{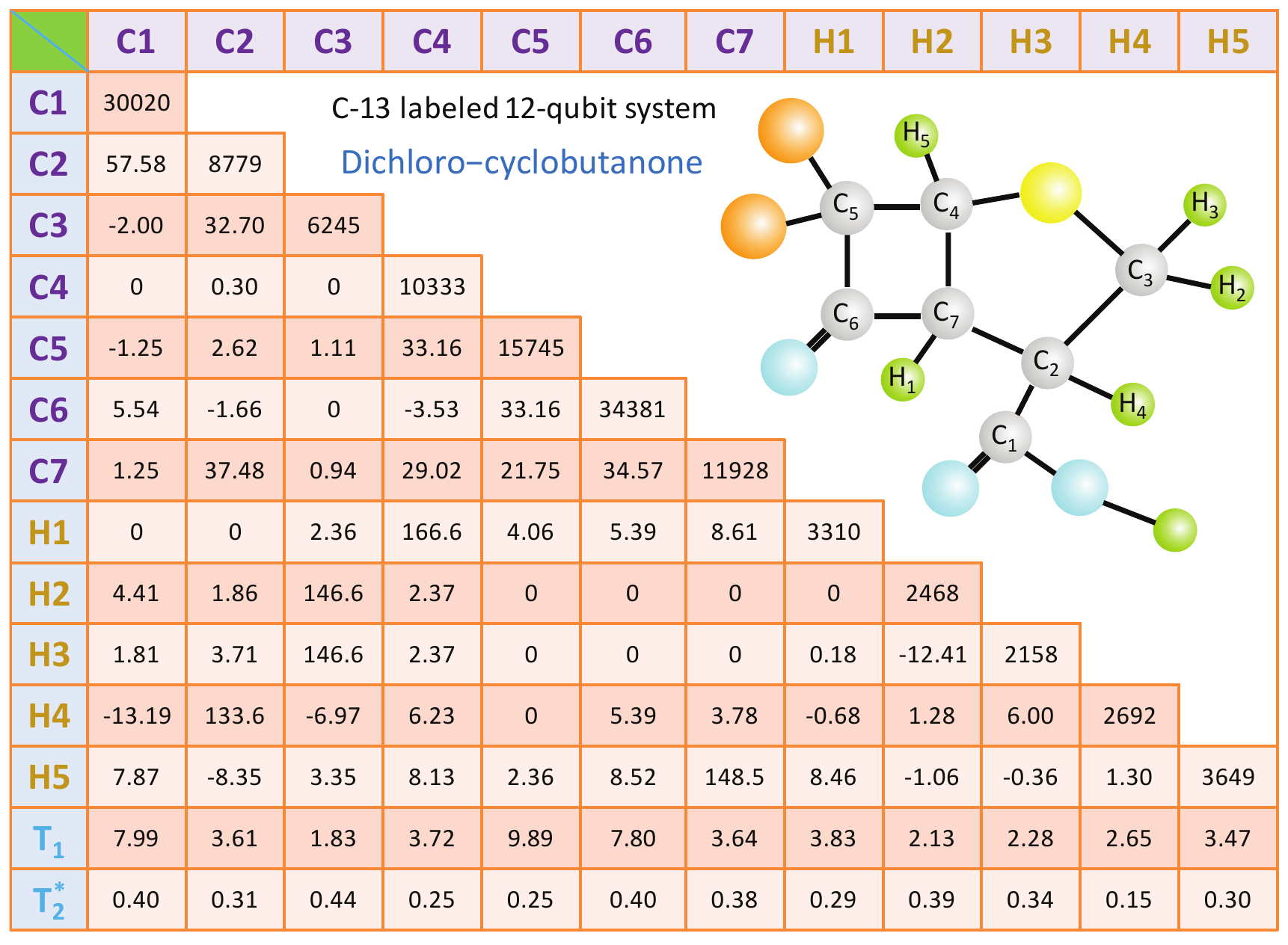}
\end{center}
\setcounter{figure}{0}
\makeatletter
\renewcommand{\thefigure}{S\@arabic\c@figure}
\makeatother
\caption{Molecular structure and Hamiltonian parameters of per-$^{13}$C labeled (1S,4S,5S)-7,7-dichloro-6-oxo-2-thiabicyclo[3.2.0]heptane-4-carboxylic acid. C$_1$ to C$_7$, and H$_1$ to H$_5$ denote the 12 qubits from qubit 1 to qubit 12, respectively. The diagonal elements are the chemical shifts (in Hz), and the off-diagonal elements are the $J$-couplings between two spins (in Hz). The relaxation times $T_{1}$ and $T_{2}$ (in seconds) are also listed at bottom.}
\label{molecule}
\end{figure*}

\appendix

\begin{figure}[t]
\begin{center}
\includegraphics[width=\linewidth]{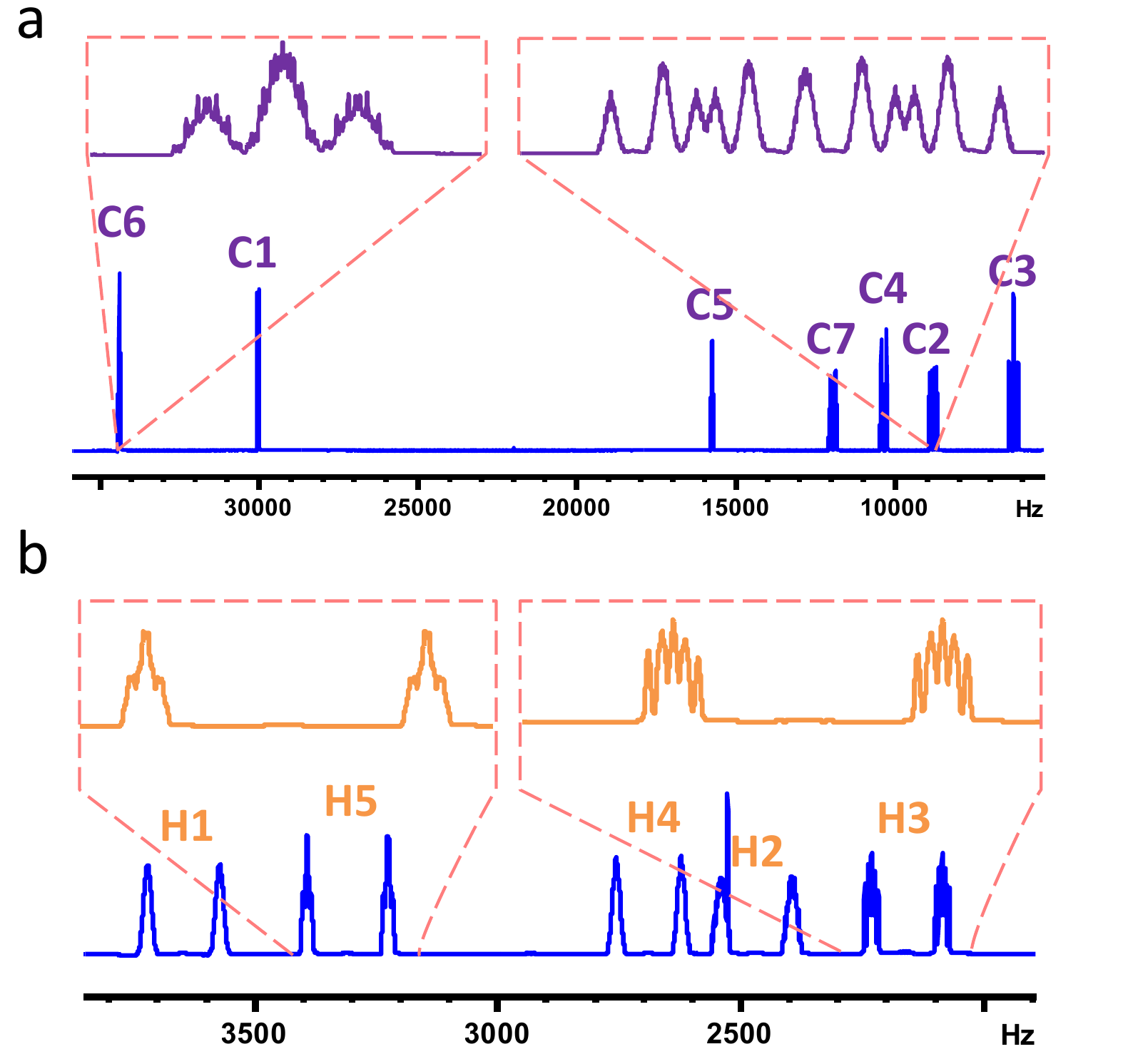}
\end{center}
\makeatletter
\renewcommand{\thefigure}{S\@arabic\c@figure}
\makeatother
\caption{Thermal spectra of \textbf{a}, $^{13}$C and \textbf{b}, $^1$H in the 12-qubit quantum processor. In particular, the spectra of C$_2$, C$_6$, H$_3$, and H$_5$ are magnified for better visualization. The $y$-axis represents the signal strength (a.u.). Different spins are individually addressed according to their distinct resonance frequencies, and the spectrum of each spin is split into up to $2^{11}$ peaks due to its couplings with the other 11 spins (though many splittings are too small to be resolved).}
\label{thermal}
\end{figure}

\section{Choice of the initial guess}

Making a good initial guess is critical for an optimization task on a large system.  For small-sized (e.g., 2-5 spins) systems, even starting from an arbitrary initial guess, it is highly possible to reach a good control solution. However, for larger quantum systems (e.g., 10 spins or more), it is not practical to start from a completely random guess.

The difficulty in choosing the initial guess is primarily due to the fact that there is no theory to tell what is the minimum time length required for a given control task. In general, for larger systems with more complex target states, we need much longer pulses for state engineering tasks. However, noticing that the slice length $\Delta t$ should be small to ensure the accuracy of gradient estimation, longer pulse leads to more time steps. A rough estimation is, for our 12-qubit system, typically the total time needed for a control pulse to prepare Z$^{12}$ from a single Z is about 100 ms. That is to say, for $\Delta t = 20$ $\mu$s (what we used in the experiment), it means 5,000 slices. Obviously, optimizing so many control variables in a $2^{12}$-dimensional space is a huge task,  in particular if one starts from a totally random guess.

A possible solution is to use the so-called sequence compiler technique \cite{ryan2008liquid} to generate a suitable initial guess. This is an efficient algorithm, and is also what we used in the current work. The basic idea is as follows.
A quantum circuit is usually constructed based on the coupling network of the controlled system. In liquid NMR, the circuit is composed of local rotations and $J$-coupling evolutions, where local rotations are generally much faster than $J$-coupling evolutions. Therefore, the dominating part of the circuit is free evolution, accompanied by some slices that correspond to local rotations. Only the local rotations are to be optimized since free evolution indicates zero control parameters. This will greatly reduce the size of the parameter space, which, for example, eventually leaves us only 110 slices to be optimized for a Z$^7$ to Z$^{12}$ preparation.

Furthermore, local operations can be optimized using compiled selective pulses as the initial guess \cite{li2016selective}. Various types of errors arise when a local rotation is realized by a selective without correction. The pulse compilation process is efficient and can eliminate the zero-th and first-order control errors, thus substantially increases the goodness of the initial guess.

\section{Optimizing Clifford gates}

While our experiment focuses on state engineering,  MQFC can also be used for other quantum optimization tasks.  As an example, we consider optimizing the pulse sequence for a generic Clifford gate.  It is possible to use twirling to estimate the average gate fidelity of a Clifford gate efficiently  \cite{lu2015experimental}.  The twirling protocol is based on finding the fidelity between  experimental states following the pulse sequence and the corresponding desired states following the ideal gate.  In principle this should be done for a complete set of initial states, but a randomized protocol can be used to approximate the gate fidelity with a constant number of experiments.  The MQFC protocol can be modified to extract the desired fidelities and optimize the pulse sequence accordingly.

For a  faulty Clifford channel $\Lambda$, its average fidelity takes the following form:
\begin{align} \label{fidelity_pr}
\bar{F}(\Lambda) = \frac{2^n \text{Pr}(0) +1}{2^n +1},
\end{align}
where $\text{Pr}(0)$ is the probability of no error. In fact, $\text{Pr}(0)$ is the linear combination of state fidelities
\begin{align} \label{noerror}
\text{Pr}(0) = \frac{1}{4^n} + \frac{1}{4^n}\sum_{k=1}^{4^n-1} \text{tr}\left( \Lambda \left( \rho_{i}^{(k)}\right) \cdot \rho_f^{(k)} \right),
\end{align}
where $\rho_{i}^{(k)} $ is one element chosen from the Pauli group and $\rho_{f}^{(k)} $ is the relevant output by applying the ideal Clifford gate on $\rho_{i}^{(k)} $. In other words, the average fidelity of a Clifford gate is a summation of   state fidelities. Although the expression of $\text{Pr}(0)$ seems to contain $4^n-1$ elements,   we can  approximately estimate its value in polynomial time by a uniform sampling approach based on the Hoeffding's inequality \cite{lu2015experimental}. In particular, we have demonstrated that one just needs to conduct a constant number of experiments ($\sim10^3$) to achieve a 99\% confidence level regardless of the system's size. Each experiment in the twirling protocol is a Pauli-to-Pauli state evolution, exactly the same as what we have done in this work. In other words, given a satisfactory confidence level, we just need to repeat MQFC for a constant number of input states, and linearly combine all fidelities and gradients of state-to-state optimizations to fulfill the task of optimizing a Clifford gate. Therefore, the complexity of optimizing a Clifford gate using MQFC is the same as that of optimizing a state-to-state evolution up to a constant, meaning that optimizing a Clifford gate via MQFC is feasible.

\section{The 12-qubit quantum processor}
Our 12-qubit processor is per-$^{13}$C labeled (1S,4S,5S)-7,7-dichloro-6-oxo-2-thiabicyclo[3.2.0]heptane-4-carboxylic acid dissolved in d6-acetone, with the molecular structure shown in Fig. \ref{molecule}.  The unlabeled compound was synthesised previously and its structure was established unambiguously by a single crystal X-ray diffraction study \cite{johnson2008cyclobutanone}.

The system Hamiltonian for the sample is given as  Eq. (1) in the main text. At room temperature, the thermal equilibrium state of this system is highly mixed, with the form
\begin{equation}
\rho_{eq}=\frac{1-\epsilon}{2^{12}} \mathbb{I} + \epsilon \left(\gamma_{\text{C}} \sum_{i=1}^7 \sigma_{z}^{i}+\gamma_{\text{H}} \sum_{j=8}^{12} \sigma_{z}^{j} \right),
\label{rhoeq}
\end{equation}
where $\epsilon \approx 10^{-5}$ describes the polarization, $ \mathbb{I}$ is a $2^{12} \times 2^{12}$ identity matrix, and $\gamma_{\text{C}}$ and $\gamma_{\text{H}}$ are the gyromagnetic ratios of the $^{13}$C and $^1$H nuclei, respectively. In particular, $\gamma_{\text{H}} \approx 4 \gamma_{\text{C}}$, so the signal of $^1$H is roughly four times as that of $^{13}$C. As the large identity part in Eq. (\ref{rhoeq}) does not contribute to the NMR spectrum under unitary evolutions, we just omit it in general and use the remaining term, the so-called \emph{deviation density matrix}, to represent the state:
\begin{equation}
\rho_{dev} \approx \sum_{i=1}^7 \sigma_{z}^{i}+ 4 \sum_{j=8}^{12} \sigma_{z}^{j},
\label{rhodev}
\end{equation}
where $\gamma_{\text{H}} \approx 4 \gamma_{\text{C}}$ is used and the polarization $\epsilon$ is dropped. In Fig. \ref{thermal}, the $^{13}$C and $^1$H thermal equilibrium spectra are shown, while the spectra of C$_2$, C$_6$, H$_3$, and H$_5$ are magnified for better visualization. All 12 spins can be individually addressed by their distinct chemical shifts as shown in Fig. \ref{molecule}. For each spin, the spectrum in principle contains $2^{11}$ peaks due to its couplings with the other 11 spins. However, many interactions, especially those between distant spins, are too small to be resolved spectrally, so the number of observable peaks is much less    than $2^{11}$.

It would be more convenient to work in the rotating frame rather than in the lab frame. For the internal Hamiltonian, we set the two transmission frequencies as $o_1 = 20,696$ Hz and $o_2 = 2,894$ Hz for the channel $^{13}$C and $^1$H, respectively. The transmission frequencies are chosen as the central frequencies of the spectra. In this double-rotating frame, the internal Hamiltonian then becomes
\begin{align}
\mathcal{H}_{s}^{rot} = {} &  -\pi\sum_{i=1}^{7} \left( \nu^i_0-o_1\right) \sigma_z^{i} -\pi\sum_{j=8}^{12} \left( \nu^j_0-o_2\right) \sigma_z^{j}  \nonumber \\
{} & + \frac{\pi}{2}\sum_{i=1<j}^{12} J_{ij} \sigma_z^{i}\sigma_z^{j}.
\label{Hrot}
\end{align}
The external control field is applied in the transverse $x$-$y$ plane,   oscillating at the transmission frequency of $^{13}$C and $^1$H channel, respectively. These transmission frequencies are in the radio-frequency (r.f.) regime. There are four control parameters, namely $\text{B}_x^\text{C}$, $\text{B}_y^\text{C}$, $\text{B}_x^\text{H}$, and $\text{B}_y^\text{H}$, in a time-independent r.f. pulse, where, without loss of generality, $\text{B}_x^\text{C}$ means the control field amplitude along $x$-axis for the $^{13}$C channel. In the double-rotating frame, the Hamiltonian of the external control field is
\be
\mathcal{H}_{c}^{rot} =  \text{B}_x^\text{C}\sum_{i=1}^7 \sigma_x^i + \text{B}_y^\text{C}\sum_{i=1}^7 \sigma_y^i+\text{B}_x^\text{H}\sum_{j=8}^{12} \sigma_x^j + \text{B}_y^\text{H}\sum_{j=8}^{12} \sigma_y^j.
\label{Hext}
\ee
Together with the internal Hamiltonian $\mathcal{H}_{s}^{rot}$ in Eq. (\ref{Hrot}), the dynamics of the total system is dominated by the joint action of the internal and external Hamiltonians, with the propagator
\be
U = e^{-i\left( \mathcal{H}_{s}^{rot}+\mathcal{H}_{c}^{rot}\right)t}.
\label{U}
\ee
All elementary gates such as single-qubit rotations and two-qubit controlled-NOT (CNOT) gates required in universal quantum information processing can be realized by deliberately designing the external Hamiltonian \cite{vandersypen2005nmr}.

For instance, if we want to realize a $\pi/2$ rotation about the $x$-axis for C$_1$, denoted as $R_x^1\left( \pi/2 \right)$, we can use a $M$-slice shaped pulse with slice width $\Delta t$. In each slice, the four parameters of the control field are constants, labeled as $\text{B}_x^\text{C}[m]$, $\text{B}_y^\text{C}[m]$, $\text{B}_x^\text{H}[m]$, and $\text{B}_y^\text{H}[m]$, respectively. The propagator of such a shaped pulse is a concatenation of the propagator in Eq. (\ref{U})
\be
U_{1}^{M} = U_MU_{M-1}\cdots U_{1},
\label{Um2}
\ee
with $U_m = e^{-i(\mathcal{H}_{s}+\mathcal{H}_{c}[m])\Delta t}$.

The next step is to find the shaped pulse, i.e., a sequence of $\text{B}_{x,y}^\text{C,H}[m]$, such that the propagator in Eq. (\ref{Um2}) realizes the target operation $R_x^1\left( \pi/2 \right)$ with high fidelity. In state-of-the-art NMR techniques, this optimization procedure is often realized via the GRAPE algorithm, as a shaped pulse found by GRAPE can have the properties of short duration and robustness to uncertainties in the Hamiltonian, e.g., the inhomogeneity of the static or control field.

\section{Subsystem-based GRAPE}

In small-scale systems with around seven spins, GRAPE is quite powerful, as it generates high-fidelity shaped pulses readily with modern computing power. However, in the 12-qubit system, GRAPE is significantly more challenging, as it requires much higher dimensional matrix multiplications and exponentiating. Therefore, we modified the original GRAPE and applied this algorithm based on subsystems, which we call subsystem-GRAPE (SSGRAPE).

We would like to stress at first that  SSGRAPE is still classical and thus cannot address the scalability issues of GRAPE \cite{ryan2008liquid}. Even though, SSGRAPE is an important modification to the original GRAPE algorithm, which can improve the timescale of calculating GRAPE pulses dramatically by defining subsystems based on the Hamiltonian of the molecule.
For example, in our 12-qubit system, by artificially disconnecting C$_2$ and   C$_7$, we divided the entire system into two subsystems with each consisting of six spins. From Fig. \ref{molecule} and the relevant parameters, it can be seen that the two subsystems are isolated to a good approximation. We define the subsystem with C$_2$ as $S_A$, and the other as $S_B$. Both internal and external Hamiltonians in $S_A$ and $S_B$ can be determined by tracing out the other subsystem. For a target operator, say $U_{tar} = R_x^1\left( \pi/2 \right)$, it can be decomposed into two operators
\be
U_{tar}^{A} = R_x^1\left( \frac{\pi}{2} \right), \quad  U_{tar}^{B} = I,
\ee
where $U_{tar}^{A}$ and $U_{tar}^{B}$ are now both $2^6 \times 2^6$ unitary operators, and $U_{tar} = U_{tar}^{A} \otimes U_{tar}^{B}$. Therefore, the 12-qubit GRAPE optimization problem can be treated as two 6-qubit problems, and SSGRAPE attempts to optimise a shaped pulse which can realize $U_{tar}^{A}$ and $U_{tar}^{B}$ simultaneously. In brief, the SSGRAPE technique greatly reduces the computation time of the pulse finding on our 12-qubit system, but it is worth emphasizing that it does not fundamentally solve the scalability issue.

%First of all, SSGRAPE is still classical, and cannot address the scalability issues of GRAPE \cite{ryan2008liquid}. Even though, SSGRAPE is an important modification to the original GRAPE algorithm, which can improve the timescale of calculating GRAPE pulses dramatically by defining subsystems based on the Hamiltonian of the molecule. For instance, in our 12-qubit molecule, we define two 6-qubit subsystems by detaching C$_2$ from C$_7$, which reduces the computational complexity greatly. Such a separation is optimal in this particular molecule, as the sole remarkable connection between the two subsystems is the coupling between C$_2$ and C$_7$. Thereafter, the total fitness function $f$ can be written as a sum of fitness functions of subsystems, as long as the target operation does not involve communications between subsystems. From Fig. \ref{fig2}(c), we know that all intended operations in our experiment are local, implying that we can apply SSGRAPE to calculate each operation in principle. Again, SSGRAPE vastly relies on the molecular structure and the operations to be optimized, leading to its scalability similar to the original GRAPE algorithm.

\begin{table}[b]
{\footnotesize
\setlength{\abovecaptionskip}{0.50cm}
\begin{tabular} {c||c|c|c|c}
\hline
  \hline
  Operator & Length & Simulated Fidelity & No. of Slices & $\Delta t$ \\
  \hline
  $\text{C}_7\text{H}_1 - \text{SWAP}$ & 8 ms & 99.0\% & 400 & 20 $\mu$s \\
  $R_y^{1-6,8-12}(\pi/2)$ & 1 ms & 99.8\% & 100 & 10 $\mu$s\\
  $R_y^7(\pi/2)$ & 1 ms & 99.9\% & 100 & 10 $\mu$s\\
  $R_x^2(\pi)$ & 2 ms & 99.8\% & 200 & 10 $\mu$s\\
  $R_x^6(\pi)$ & 2 ms & 99.8\% & 200 & 10 $\mu$s\\
  $R_x^4(\pi)$ & 2 ms & 99.7\% & 200 & 10 $\mu$s\\
  $R_x^{5,7}(\pi)$ & 2 ms & 99.8\% & 200 & 10 $\mu$s\\
  $R_{-y}^7(\pi/2)$ & 1 ms & 99.9\% & 100 & 10 $\mu$s\\
  $R_y^2(\pi/2)$ & 1 ms & 99.9\% & 100 & 10 $\mu$s\\
  $R_x^1(\pi)$ & 2 ms & 99.8\% & 200 & 10 $\mu$s\\
  $R_x^{2,3}(\pi)$ & 2 ms & 99.8\% & 200 & 10 $\mu$s\\
  $R_{-y}^2(\pi/2)$ & 1 ms & 99.9\% & 100 & 10 $\mu$s\\
  \hline
  \hline
\end{tabular}
}
\caption{Shaped pulse optimized by SSGRAPE during the 12-coherence creation. The pulses are listed in the order of their appearances in Fig. 2(c). Although the pulses are found with the subsystem method, the fidelities reported here are calculated on the full 12-qubit system.}
\label{tableop}
\end{table}

Another requirement of adopting SSGRAPE is that the target unitary operator can be effectively decomposed using subsystems and does not involve interactions between subsystems. In our 12-qubit experiment, this condition holds for every operator. We list all the SSGRAPE-optimized shaped pulses that are needed in the experiment, as shown in Table \ref{tableop}. We also simulated the fidelity of each pulse in the full 12-qubit system. That is, each pulse was found using SSGRAPE in the two 6-qubit subsystems, but then simulated on the full system. All local pulses are over $99.7\%$ fidelity in simulation, which demonstrates that SSGRAPE is a valid pulse searching method for our 12-qubit system.

\section{Experimental implementation of creating a 7-coherence}
In this section, we present a step-by-step description of our experiment of creating the 7-coherence $\text{Z}^{\otimes 7}\text{I}^{\otimes 5}$, and show the relevant NMR spectra at each step.

%\begin{figure*}
%\begin{center}
%%\includegraphics[width=0.5\linewidth,angle=-90]{circuit.pdf}
%\includegraphics[width=0.5\linewidth]{circuit.pdf}
%\end{center}
%\caption{Quantum circuit that evolves the system from thermal equilibrium state to 12-coherence. Four intermediate states $\rho_a$ to $\rho_d$ are marked at the top, which assists in the understanding of the entire circuit. In particular, the MQFC procedure is applied on 7-coherence Z$^{\otimes 7}$I$^{\otimes 5}$.}
%\label{circuit}
%\end{figure*}

\subsection{From the initial state to $\rho_a$}

The whole circuit is depicted in Fig. 2(c) in the main text, with four intermediate states labeled by $\rho_a$, $\rho_b$, $\rho_c$, and $\rho_d$. The initial state is  the thermal equilibrium state: $\rho_{dev} = \sum_{i=1}^7 \text{Z}_{i}+ 4 \sum_{j=8}^{12} \text{Z}_{j}$, as was described in Eq. (\ref{rhodev}). It first undergoes an 8 ms SWAP gate, which swaps the equilibrium polarizations between C$_7$ and H$_1$. In   doing so,   the signal of C$_7$ is boosted by approximately four times. After this, a multi-qubit rotation about the $y$-axis on all spins except C$_7$ is applied to rotate these spins to the transverse plane, followed by a $z$-direction gradient field. A $z$-gradient pulse is used to destroy non-zero coherences, i.e., removes all the Pauli terms that contain X and Y terms in our case. As all the other spins except C$_7$ are flipped to the $x$-$y$ plane, the resulting state after the gradient pulse is
\be
\rho_a = \text{Z}_7.
\label{rhoa}
\ee
Here, we have ignored the factor of four before $\text{Z}_7$ for convenience, as this state will be used as the reference for later calibrations. When we observed C$_7$ by rotating it to X, the two spectra of $\rho_a$ are shown in Figs. \ref{spec1}(a) and \ref{spec1}(b). The left one is the spectrum of C$_7$ in the 12-qubit regime, and the right one is obtained by decoupling the $^1$H channel via the Waltz-16 sequence \cite{shaka1983evaluation}. This decoupling can be considered as a partial trace process, which can remarkably improve the spectrum resolution, but it requires that the state of the five $^1$H nuclei is equal to the identity.

\subsection{From $\rho_a$ to $\rho_b$}

The next step is to create a 5-coherence on the nearest neighbours of C$_7$, including C$_2$, C$_4$, C$_5$, and C$_6$. Let us start from a simple example to describe how to increase the coherence order. For two qubits, if we start from XI, choose $t = 1/(2J)$, and let the system evolve under the $J$-coupling of $\sigma_z \sigma_z$ term, the coherence order of the system can be increased by one according to
\be
\text{XI} \xrightarrow{U\left( 1/2J\right) =  \pi \sigma_z^1 \sigma_z^2/4} \text{YZ}.
\label{zz}
\ee
The main idea of creating 5-coherence is to make use of the partial refocusing scheme \cite{lu2015experimental}, that only the desired $J$-coupling evolutions are left to undergo $t = 1/(2J)$ evolutions, while all the unwanted couplings are refocused. Refocusing of an unwanted $\sigma_z \sigma_z$ coupling term can be realised by inserting a $\pi$ pulse on one spin in the centre of the evolution but no pulse on the other. Although the desired couplings $J_{27}$, $J_{47}$, $J_{57}$, and $J_{67}$ are different, a simultaneous $J$-coupling evolution is possible through careful design of the $\pi$-pulses' positions; see Fig. 2(c).

\begin{figure*}
\begin{center}
\includegraphics[width= \linewidth]{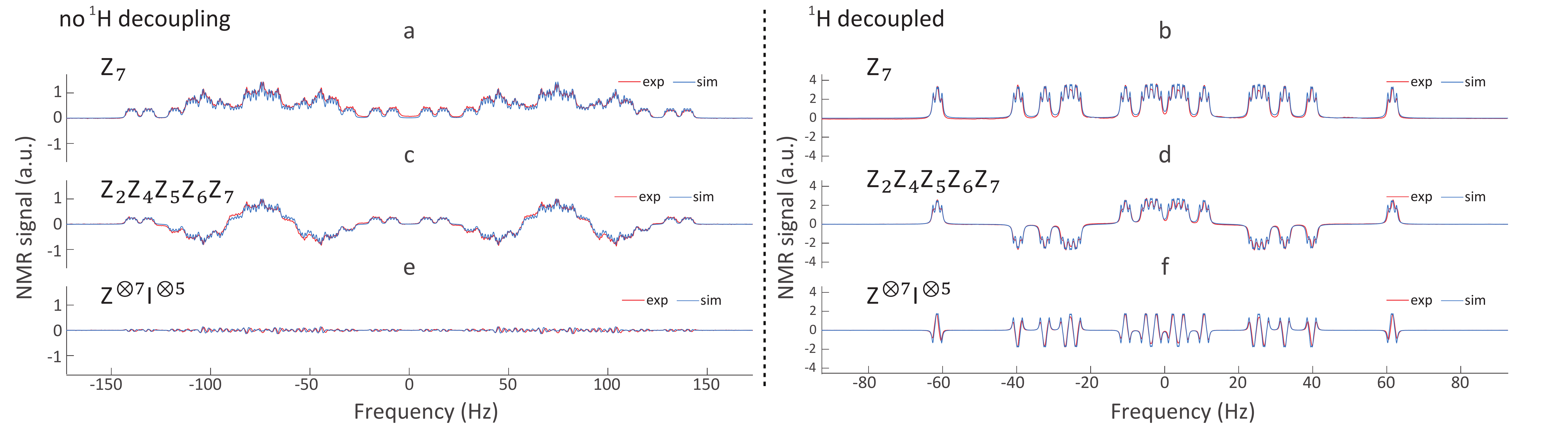}
\end{center}
\makeatletter
\renewcommand{\thefigure}{S\@arabic\c@figure}
\makeatother
\caption{Spectra for the observation of $\rho_a$ (\textbf{a}, \textbf{b}), $\rho_b$ (\textbf{c}, \textbf{d}), and $\rho_c$ (\textbf{e}, \textbf{f}) on C$_7$, respectively. The left column is without $^1$H decoupling, and the right column is with $^1$H decoupled. The spectra in the left column are averaged over 30 scans to gain a good signal-to-noise ratio, while the ones in the right column are averaged over only 10 scans. In each spectrum, the experiment is in strong agreement with simulation, indicating that our control on this 12-qubit system is precise. }
\label{spec1}
\end{figure*}

After the partial refocusing sequence and a subsequent $R_y^7\left( -\pi/2 \right)$ pulse that rotates C$_7$ back to Z, the ideal state at point $b$ is
\be
\rho_b = \text{Z}_2\text{Z}_4\text{Z}_5\text{Z}_6\text{Z}_7.
\label{rhob}
\ee
In experiment, we observed C$_7$ for this 5-coherence state, and the two spectra without and with decoupling the $^1$H channel are shown in Figs. \ref{spec1}(c) and \ref{spec1}(d), respectively. The signal attenuation in experiment is about 20.7\% due to decoherence, so the simulated spectra were rescaled by 1.26 times (compared to the simulated spectra of $\rho_a$) to fit the experimental data.

\subsection{From $\rho_b$ to $\rho_c$}

In this step, we create the 7-coherence involving all the $^{13}$C nuclei. Coherence is transferred to the remaining C$_1$ and C$_3$ spins from their joint neighbour C$_2$. Similar to the above procedure, this step also involves a partial refocusing sequence, which realises the $t = 1/(2J)$ evolutions for $J_{12}$ and $J_{23}$ simultaneously. After a local pulse $R_y^2\left( -\pi/2 \right)$ on C$_2$, the state at point $c$ is
\be
\rho_c = \text{Z}^{\otimes 7}\text{I}^{\otimes 5},
\label{rhoc}
\ee
where $\text{I}^{\otimes 5}$ indicates that all five $^1$H's are still in the identity state. The experimental spectra of C$_7$ without and with $^1$H decoupling are plotted in Figs. \ref{spec1}(e) and \ref{spec1}(f), respectively. The simulated spectra were rescaled by 1.42 (compared to the simulated spectra of $\rho_a$) to make an optimal fit with the experimental result.

Despite the non-negligible decoherence effect during the experimental creation of 7-coherence, we emphasize that this signal attenuation will not impact the characterization of the MQFC procedure, which is the main focus of this work. As shown in Figs. \ref{spec1}(e) and \ref{spec1}(f), the creation of 7-coherence is remarkably precise up to a rescaling factor. This 7-coherence state can be used as a reference to calibrate the 12-coherence created via MQFC. Hence, the scaling factor of the 7-coherence state is irrelevant. We also direct readers to Ref. \cite{lu2015experimental} for a detailed calibration of our 7-coherence result and the relevant spectrum of another spin C$_2$.

\section{E\lowercase{xperimental} MQFC \lowercase{optimization to create 12-coherence}} \label{12coh}

\begin{figure*}
\begin{center}
\includegraphics[width= \linewidth]{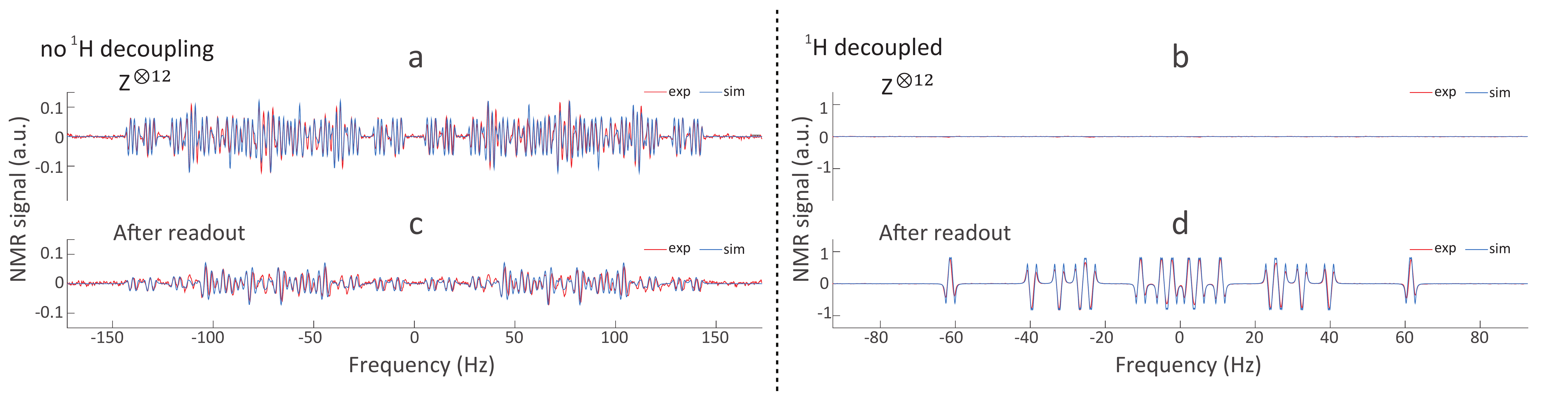}
\end{center}
\makeatletter
\renewcommand{\thefigure}{S\@arabic\c@figure}
\makeatother
\caption{Spectra for the observation of 12-coherence $\rho_d$ (\textbf{a}, \textbf{b}) and after the readout stage (\textbf{c}, \textbf{d}) on C$_7$, respectively. The left column is without $^1$H decoupling, and the right column is with $^1$H decoupled. The $^1$H undecoupled spectra are averaged over 30 scans. As shown in \textbf{a}, the direct observation of experimental 12-coherence $\rho_d = \text{Z}^{\otimes 12}$ matches remarkably well with the simulation, which demonstrates that we have successfully created 12-coherence using the MQFC pulse. Not surprisingly, the decoupling of $^1$H leads to no signal in \textbf{b}, since the five $^1$H's are no longer identity but $\text{Z}^{\otimes 5}$. \textbf{c} and \textbf{d} are the spectra after the readout stage in Fig. \ref{fig2}(c), whose purpose is to measure $f$ and $g$ in one scan per experiment. }
\label{spec2}
\end{figure*}

The central focus of this work is to  design  a shaped pulse based on MQFC to create 12-coherence from 7-coherence. More precisely, in the circuit of Fig. 2(c),   we want to optimize  the part between $\rho_c$ and $\rho_d$. Unlike the preceding section where all pulses were calculated by   SSGRAPE,   MQFC optimization is   quantum, as the fitness function $f$ and gradient $g$ were directly measured on the 12-qubit quantum computer. The only role that a classical computer played during MQFC was to update the control field in terms of the measured gradient $g$, which is merely simple algebra without additional time cost. That is, no inefficient calculations on classical computers were involved during MQFC.

After the creation of 7-coherence $\rho_c= \text{Z}^{\otimes 7}\text{I}^{\otimes 5}$, we applied a $z$-direction gradient field as depicted in Fig. 2(c).  The purpose is to remove unwanted terms produced due to experimental imperfections, since $\text{Z}^{\otimes 7}\text{I}^{\otimes 5}$ itself is invariant under this gradient field. This technique is conventional in NMR quantum computing   to `clean up' the experimentally prepared input state, and has no influence on the subsequent MQFC procedure.

The structure of the shaped pulse used in MQFC was pre-designed according to the molecular information in Fig. \ref{molecule}. It consists of five parts: three sub-pulses and two free evolutions in between sub-pulses as shown in the \emph{MQFC optimization} box in Fig. 2(c). The general idea of this structure design is to let $^{13}$C's interact with five $^1$H's simultaneously and hence increase the coherence order by five. Note that all large C-H couplings in Fig. \ref{molecule} are   roughly $J_{ave}=148.8$ Hz on average. Therefore, we set the time for the two free evolutions as $1/4J_{ave} \approx 1.68$ ms, and expect that it enables sufficient C-H interaction time to produce higher coherence on the five $^1$H's. The functions of   the three sub-pulses are: the first one is to rotate C$_2$, C$_3$, C$_4$, and C$_7$, which are directly connected to $^1$H, to the $x$-$y$ plane; the second sub-pulse is to refocus unwanted couplings during the C-H interaction; the last one is to rotate the relevant spins back to Z. The MQFC pulse is set to be 5.56 ms  with $\Delta t=20$ $\mu$s. The total number of slices is thus 278, where 168 of them remain zero as they are meant for free evolutions. The remaining 110 slices are divided into three parts: 30 for the first sub-pulse, 30 for the second sub-pulse, and 40 for the third sub-pulse. Hence, to measure the gradient $g$ in each iteration, we only need to take these 110 slices into account, which greatly reduces the experimental running time.

Since MQFC is a gradient-based optimization procedure, the measurements of $f$ and $g$ are critical. As explained before, measuring $f$ is actually equivalent to measuring the expectation value $\langle \text{Z}^{\otimes 12} \rangle$ in the experimental state after applying the trial shaped pulse. This measurement requires only one experiment. Analogously, measuring $g$ also involves the readout of the expectation value $\langle \text{Z}^{\otimes 12} \rangle$, with a $\pi/2$ local pulse inserted in the trial shaped pulse (see Fig. 2(b)  and Eq. (11)). This measurement requires $4nM$ experiments, where $n=7$ because we only need to apply local $\pi/2$ pulses on the seven $^{13}$C's, and $M = 110$ is the number of slices as described in the preceding paragraph. In total, for each iteration, the experimental time of MQFC is
\be
T_{exp}=  (4nM+1)\tau_{exp},
\label{time}
\ee
where $\tau_{exp}$ is dominated by the delay time between two experiments to reestablish thermal equilibrium. Typically, $\tau_{exp} \approx 5T_1$, implying a 30 s delay between experiments. However, the observation of $\text{Z}^{\otimes 12}$ in our 12-qubit system requires about 30 experimental scans to yield a good spectrum with acceptable signal-to-noise ratio (SNR), such as the one in Fig. \ref{spec2}(a). Estimated by Eq. (\ref{time}), this requirement leads to a time cost of over one month per iteration, which is impractical as in this experiment we used nine iterations to achieve a high-fidelity MQFC pulse.

Therefore, we need to improve the SNR of the spectrum in order to reduce the number of scans and thus reduce the experimental time. A traditional way is to decouple $^1$H spins, which should enhance the SNR by $2^5=32$ times, because the NMR signal per peak attenuates exponentially with the number of interacting spins. However, when the five $^1$H's are in $\text{Z}^{\otimes 5}$, the decoupling, which in fact traces out $^1$H, would lead to no signal on $^{13}$C as shown in Fig. \ref{spec2}(b). In other words, it is necessary to evolve the state of $^1$H to $\text{I}^{\otimes 5}$ before decoupling. In the experiment, we used a readout pulse to realize this transformation.

\section{Readout sequence for the measurement of $f$ and $g$ }\label{readout2}

As mentioned above, the direct observation of 12-coherence $\text{Z}^{\otimes 12}$ requires about 30 scans to yield a good SNR in the spectrum. Compared to the $^1$H decoupled spectrum which merely requires one scan, the experimental time of the undecoupled case is 30 times longer and thus impractical for measuring $f$ and $g$. This section is to describe our readout technique, which enables the decoupling of $^1$H's so that each experiment can be done with only one scan.

A readout sequence (see Fig. 4), computed by classical SSGRAPE, is run just after the MQFC procedure. The phase correction is a $z$-rotation to compensate for the unwanted chemical shift evolutions during $1/2J_{78}$ time. If the state is $\text{Z}^{\otimes 12}$, the five $^1$H's will   evolve to the identity state after the readout sequence, and the decoupling of $^1$H will lead to the C$_7$ spectrum in Fig. \ref{spec2}(d), which can be measured with a single scan. We used Lorentzian fitting to obtain the signal's amplitude and phase, and thus the value of $\langle \text{Z}^{\otimes 12} \rangle$.

This readout sequence would inevitably induce errors due to  the decoherence and pulse imperfections. For the former error source, through our simulation we found that the readout caused about $30\%$ signal loss, which is reasonable since multi-coherence is exceptionally vulnerable to decoherence. Therefore, this factor was taken into account for all the measurement results, that is, the measured values are rescaled by 1.3. As to the pulse imperfection, it consists of two parts: the imperfection of the sequence itself, i.e. some approximations about $J$-couplings when designing this simple readout sequence, and the infidelity of the SSGRAPE pulse. In total, 3.5\% error arises in simulation, but how the error affects the 12-qubit quantum states is difficult to quantify. We used this value as the uncertainty of the experimental value of $\langle \text{Z}^{\otimes 12} \rangle$, namely, the error bars in Fig. 3c in the main text. Fortunately, MQFC outperforms SSGRAPE, even with error bars accounted for, demonstrating that MQFC has the feedback-control property that is able to correct unknown experimental errors.

\begin{figure}[t]
\begin{center}
\includegraphics[width=\linewidth]{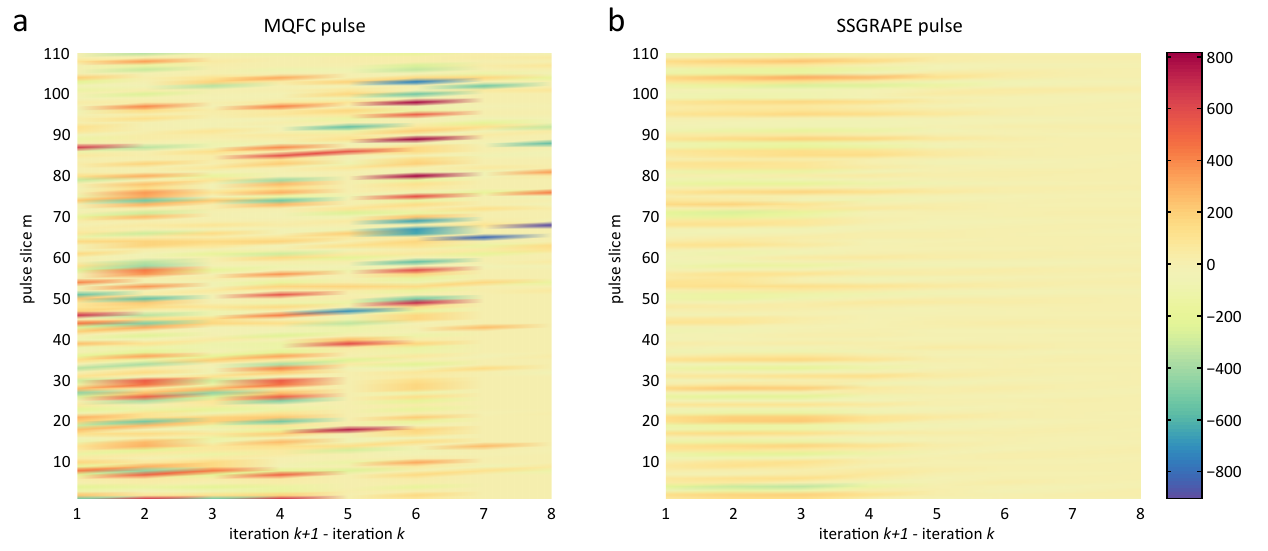}
\end{center}
\makeatletter
\renewcommand{\thefigure}{S\@arabic\c@figure}
\makeatother
\caption{Variations of $\text{B}_x$ between iterations $k+1$ and $k$, indicated by the $x$-axis, for \textbf{a}, MQFC pulse and \textbf{b}, SSGRAPE pulse. The $y$-axis represents the 110 slices in optimization (see Section \ref{12coh}), and $\Delta B_x$ is plotted in colourscale. Ideally, the two plots should be the same, as MQFC is measuring the gradient information on a quantum computer, which should not be different from the classical SSGRAPE calculations. Experimentally, however, they are seen to differ. This reflects the unknowns in the experimental system (uncertainties in the Hamiltonian, control fields, etc.), and demonstrates that MQFC is able to correct for these unknowns. }
\label{gradient}
\end{figure}

In addition, we plotted the variations of $\text{B}_x$ between two iterations for both MQFC and SSGRAPE pulses in Fig. \ref{gradient}. $\Delta \text{B}_x$ is proportional to the measured $g_x$, and note the $\Delta t = 20$ $\mu$s factor in the form of $g_x$ in Eq. (11), computed by $\Delta \text{B}_x = \epsilon g_x$, where $\epsilon$ is a fixed step size. In experiment, we chose $\epsilon = 1.6\text{e}^7$ according to the knowledge gained in SSGRAPE calculation. Note that $\epsilon$ can also be efficiently altered using a quadratic fit process, which is a potential improvement of the current experiment by speeding up the convergence of the optimization procedure. The difference of $\Delta \text{B}_x$ in Fig. \ref{gradient}(a) and \ref{gradient}(b) reflects that unknowns in the experimental system (uncertainties in the Hamiltonian, control fields, etc.) are automatically accounted for by MQFC, confirming its feedback control property.

It is worth stressing that the readout technique used in our experiment is merely to reduce the time cost in measuring $f$ and $g$. For other systems in which the signal is not exponentially decreased with the growing number of qubits, this readout stage is not necessary. Even in NMR, if we can shorten the reset time between two experiments, a greater number of scans can be done. Preliminary progress has been made towards this goal in our recent work \cite{li2014approximation}.

\end{document}